\documentclass[journal]{IEEEtran}
\usepackage{cite}
\usepackage{algorithmic}
\usepackage{graphicx}
\usepackage{textcomp}
\usepackage{cite}
\usepackage{amsmath,amssymb,amsfonts}
\usepackage{algorithmic}
\usepackage{algorithm}
\usepackage{graphicx}
\usepackage{textcomp}

\usepackage{amsthm}
\usepackage{xcolor}

\let\labelindent\relax
\usepackage[inline]{enumitem}

\makeatletter \def\epstopdf@sys@cmd{epstopdf} \makeatother
\def\BibTeX{{\rm B\kern-.05em{\sc i\kern-.025em b}\kern-.08em
    T\kern-.1667em\lower.7ex\hbox{E}\kern-.125emX}}
\begin{document}
\title{Phase-Space Function Recovery for Moving
Target Imaging in SAR by Convex Optimization}
\author{Sean Thammakhoune, Bariscan Yonel \IEEEmembership{Member, IEEE}, Eric Mason, Birsen Yazici \IEEEmembership{Fellow, IEEE}, Yonina C. Eldar, \IEEEmembership{Fellow, IEEE}
\thanks{Manuscript submitted April 15, 2021;revised August 7th, 2021;accepted August 9th, 2021. This work was supported in part
by the Air Force Office of Scientific Research (AFOSR) under the agreement
FA9550-19-1-0284, in part by Office of Naval Research (ONR) under the
agreement N0001418-1-2068, in part by the United States Naval Research Laboratory (NRL) under the agreement N00173-21-1-G007, and in part by the National Science Foundation
(NSF) under Grant ECCS-1809234. (Corresponding author: Birsen Yazici.) }
\thanks{The authors Sean Thammakhoune, Bariscan Yonel, and Birsen Yazici are with the Department of Electrical, Computer, and Systems
Engineering, Rensselaer Polytechnic Institute, Troy NY, 12180 USA (e-mail:
thamms@rpi.edu; yonelb2@rpi.edu; yazici@ecse.rpi.edu).}
\thanks{Eric Mason is with the United States Naval Research Laboratory, 4555 Overlook Ave SW, Washington, DC 20375 USA (e-mail: eric.mason@nrl.navy.mil)}
\thanks{Yonina C. Eldar is with the Department of Mathematics and Computer Science, Weizmann Institute of Science, Rehovot 7610001, Israel (e-mail: yonina.eldar@weizmann.ac.il)}}

\maketitle

\begin{abstract}
In this paper, we present an approach for ground moving target imaging (GMTI) and velocity recovery using synthetic aperture radar. We formulate the GMTI problem as the recovery of a phase-space reflectivity (PSR) function which represents the strengths and velocities of the scatterers in a scene of interest. We show that the discretized PSR matrix can be decomposed into a rank-one, and a highly sparse component corresponding to the stationary and moving scatterers, respectively. We then recover the two distinct components by solving a constrained optimization problem that admits computationally efficient convex solvers within the proximal gradient descent and alternating direction method of multipliers frameworks. Using the structural properties of the PSR matrix, we alleviate the computationally expensive steps associated with rank-constraints, such as singular value thresholding. Our optimization-based approach has several advantages over state-of-the-art GMTI methods, including computational efficiency, applicability to dense target environments, and arbitrary imaging configurations. We present extensive simulations to assess the robustness of our approach to both additive noise and clutter, with increasing number of moving targets. We show that both solvers perform well in dense moving target environments, and low-signal-to-clutter ratios without the need for additional clutter suppression techniques.
\end{abstract}


\section{Introduction}
\subsection{Motivation}
Synthetic aperture radar (SAR) is a well-established imaging modality for defense and remote sensing applications \cite{moreira2013tutorial,chen2017overview,ertin2007gotcha}.
Conventional SAR image reconstruction algorithms rely on the assumption of the stationarity of scatterers to be imaged. Fundamentally, imaging of moving scatterers poses a significant challenge in SAR. Moving scatterers induce velocity-dependent range variations at the antenna look-directions, resulting in artifacts in the reconstructed imagery \cite{duman2015moving}. If the velocities are known, imaging algorithms can be easily modified to compensate for the range variations. However, since motion parameters are typically unknown, moving scatterers appear mispositioned and smeared.
Ground moving target imaging (GMTI) therefore remains an important problem in a number of SAR modalities and requires additional considerations in image reconstruction and system design \cite{wang2011doppler,duman2015moving,son2016passive,wacks2014passive,
wacks2014passive_part2,wang2014bistatic,wacks2017doppler,
wang2012passive,wang2012passive2,quegan1997spotlight,
franceschetti1999synthetic,jakowatz2012spotlight,wang2013ground}.

In this paper, we present a framework for reconstructing focused images and recovering their two-dimensional velocities based on the \emph{phase-space reflectivity} (PSR) of the scene of interest, which is a function that depends on two-dimensional space and velocity variables \cite{wang2014bistatic,duman2013ground,wacks2014passive,wacks2014passive_part2,wang2012passive2}.
We formulate SAR GMTI as an optimization problem over the discretized PSR matrix, utilizing its special mathematical structure. Namely, we show that the PSR matrix is a superposition of two matrices belonging to complementary orthogonal subspaces. One of these matrices is rank-one corresponding to stationary scatterers, and the other is a highly sparse matrix corresponding to moving targets. Taking into account this structure, we obtain a constrained minimization problem that is convex over each component, and design computationally efficient solvers. We then reconstruct the stationary and moving targets simultaneously over the imaging grid along with their corresponding velocities by recovering the components of the PSR matrix.

Our approach requires only a single receiver channel, with no particular specifications on the imaging geometry or flight trajectories. It does not require prior knowledge on the number of moving targets or their sparsity in the scene, as the PSR matrix is inherently sparse by design. Additionally, we design the PSR matrix such that it admits simple solvers commonly associated with well-known sparse recovery problems. While we present our method for monostatic SAR, it is also applicable to MTI using bi-static and multi-static radar, as well as sonar.
\subsection{Related Work}
GMTI is conventionally addressed using two or more antennas traversing the same trajectory in order to provide a frame of reference to detect and separate the moving targets from the stationary background, or clutter.
This spatial diversity forms the main principle in classical space-time processing approaches such as the displaced phase center antenna, and along track interferometry \cite{muehe2000displaced,gierull2003raw,budillon2008amplitude}. 
Notably, while computationally efficient, these techniques require a specific imaging geometry with identical antennas, which increases operational costs and acquisition complexity. On the other hand, while alleviating the reliance on multiple receiver channels,  space-time adaptive processing techniques construct a two-dimensional space-time filter that requires the inversion of a large covariance matrix, which is computationally expensive to compute \cite{pettersson2004detection,ender1999space,ender1996detection,chen2004performance}. 
In addition, target-free training samples are needed to construct the filter, which can be difficult when moving target parameters are unknown.

In recent years, there has been growing interest in SAR GMTI using a single receiver together with iterative optimization methods. These are spearheaded by sparsity-driven techniques to focus moving targets in the reconstructed imagery through velocity estimation and phase correction \cite{onhon2011sar,onhon2017sar,chen2017refocusing,8867942,7422105,ccetin2014sparsity}.
In \cite{onhon2011sar}, a three-step nonquadratic regularization approach is used to correct for the phase error introduced by moving targets. In \cite{onhon2017sar}, a method that simultaneously optimizes over both the reflectivities and phases is used. In \cite{chen2017refocusing}, a parametric sparse representation is proposed to refocus moving targets. In \cite{8867942}, the received SAR signal is parameterized using a polynomial basis, allowing the effect of moving targets to become distinguishable via a sparse representation. Notably, \cite{onhon2011sar,onhon2017sar,chen2017refocusing,8867942,7422105,ccetin2014sparsity} require relatively high signal-to-clutter ratio (SCR) for adequate performance, while the parametric representations \cite{chen2017refocusing,8867942} rely on region-of-interest data, along with additional use of detection algorithms. 

Another class of GMTI techniques can be categorized as robust principal component analysis (RPCA)-based methods\cite{yang2015strong,li2016moving,schwartz2020change,xu2019joint,leibovich2019low,jia2019clutter,Yan2013,Ender,Yasin2017,Huang2020,Cetin2018,Biondi2016,Borcea2012}. In contrast to sparsity-driven techniques, this approach excels in separating the scene into stationary and moving target components, under the assumption that these components exhibit low-rank and sparse properties, respectively. 
A popular approach satisfying these assumptions is based on multi-channel SAR, which uses an antenna array of receivers \cite{yang2015strong,li2016moving,schwartz2020change,xu2019joint,leibovich2019low,jia2019clutter,Yan2013,Ender,Yasin2017,Huang2020,Cetin2018,Biondi2016}. In \cite{leibovich2019low} optimal parameters for the RPCA algorithm are derived, showing that the separation is dependent on the nuclear-norm of the sparse part related to target velocity. In \cite{jia2019clutter}, RPCA is used to determine target-free training samples used for space-time adaptive processing to improve clutter suppression. However, this method requires the construction of a computationally expensive filter. Alternatively, in \cite{Yasin2017,Huang2020,Cetin2018,Biondi2016,Borcea2012}, the monostatic SAR data matrix is decomposed into subapertures to produce low-rank characteristics for the stationary background. In \cite{Yasin2017,Cetin2018}, a full-resolution image is reconstructed by choosing non-overlapping windows for the subaperture decomposition. While RPCA-based methods separate the received signal data scattered back from stationary and moving scatterers, they require additional iterative algorithms to estimate motion parameters.
\subsection{Advantages of Our Work}
{Our formulation differs from other sparsity-driven techniques or RPCA-based methods by taking advantage of the particular structure of the PSR matrix to introduce additional constraints to the optimization problem. This vastly improves the complexity of the iterative solvers originally designed for RPCA, as it eliminates the expensive singular value thresholding step. This improvement is a direct result of the conceptually simpler structure of the problem formulation than that of \cite{yang2015strong,li2016moving,schwartz2020change,xu2019joint,leibovich2019low,jia2019clutter,Yan2013,Ender,Yasin2017,Huang2020,Cetin2018,Biondi2016,Borcea2012}. 
Unlike the generic RPCA formulation, we do not pursue a low-rank representation in the received data, or the image domain as obtained by multi-channel or sub-aperture processing, but formulate the problem over the PSR matrix that naturally admits its particular structure. Hence, we do not need to estimate the subspace that the rank-one component of the PSR matrix resides in, and a simple projection suffices to enforce the structure of the known support for the stationary scene indicated by the zero-velocity index within the PSR matrix. In other words, the nuclear-norm in the generic RPCA optimization problem formulation is replaced by an indicator function. 
Although in general rank-constraints introduce non-convexity, since the right singular vector of the rank-one component is fixed in the optimization, our formulation remains convex over both sub-problems in each variable.

{In addition, our formulation provides several advantages over the state-of-the-art in SAR GMTI: \begin{enumerate*}[label=(\roman*)] \item{Unlike\cite{onhon2011sar,onhon2017sar,ccetin2014sparsity}, it is capable of simultaneously recovering stationary and moving targets without the need for additional clutter suppression techniques. This is a direct consequence of utilizing the inherent structure of the PSR matrix.}\item{Our method also provides the two-dimensional velocities of moving targets regardless of their speed, direction, or location.} \item{While we use a range of velocities in recovering the PSR matrix, our method does not require a priori knowledge of target motion parameters.}
\item{Another key advantage of our approach is its capability to image dense target environments, because the moving target component of the PSR matrix has structural sparsity without an explicit assumption on the number of moving targets present in the scene.}\item{Unlike \cite{Yasin2017,Cetin2018} that rely on low-resolution subaperture images, our approach utilizes full-resolution images, thereby improving robustness in low SCR.}\item{Our reconstruction method does not rely on specific assumptions on the forward model or imaging configurations. This promotes the applicability of our approach to arbitrary imaging geometries and other SAR modalities.} \item{Our method reduces the implementation costs required by conventional space-time processing techniques.}\end{enumerate*}

We demonstrate these advantages over state-of-the-art methods and verify that our approach is robust to additive noise and clutter at low-SCR through extensive numerical simulations using proximal gradient descent (PGD) derived specifically for the problem formulation. We also explore the viability of alternating direction method of multipliers (ADMM), which has empirically shown improved convergence speeds over PGD. To provide a further comparison to techniques such as \cite{li2016moving}, we also derive a non-convex algorithm that assumes the number of moving targets is known a priori.

\subsection{Organization of the Paper}
The rest of the paper is organized as follows: in Section \ref{Forward Models}, we present the SAR received signal model. In Section \ref{algorithms}, we introduce our problem formulation and optimization framework for reconstructing the reflectivities of moving targets and the stationary scene. Section \ref{Numerical Experiments} provides numerical simulations results, and Section \ref{Conclusion} concludes the paper. 
\section{Received Signal Model} \label{Forward Models}
Throughout the paper, we use bold upper-cased fonts to denote matrices, bold lower-cased non-italic fonts to denote vectors in 3D, bold lower-cased italic fonts to denote 2D vectors, and non-bold lower-cased italic fonts to denote scalar quantities, i.e.,  $\mathbf{X} \in \mathbb{R}^{M\times N},\ \mathbf{x} = [x_1,\ x_2,\ x_3]\in \mathbb{R}^3,\ \pmb{x} = [x_1,\ x_2]\in\mathbb{R}^2,\ x_i \in \mathbb{R},\ i=1,2,3.$ We will use calligraphic letters, such as $\mathcal{F}$ for operators.
\subsection{SAR Forward Model}
We consider a monostatic SAR system in which the transmitter and receiver are collocated; a typical configuration is shown in Fig. \ref{mono_geometry}.
\begin{figure}[!htb]
\centering
\includegraphics[width = \linewidth]{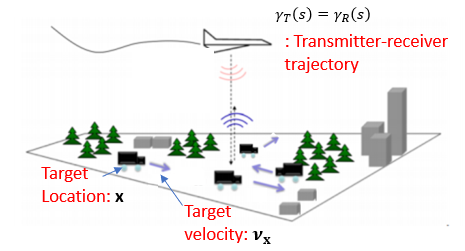}
\caption{Typical Geometry for Monostatic SAR.}
\label{mono_geometry}
\end{figure}
We begin by making the start-stop approximation in which the scatterers and the antennas only move between pulses, and are stationary during a pulse. Let $s \in [s_0,s_1]$ denote the slow-time, which indexes each data processing window and \textbf{x} be a location on the ground where \textbf{x} = [$\textit{\textbf{x}}, \psi(\textit{\textbf{x}})$] $\in \mathbb{R}^3$, $\textit{\textbf{x}} \in \mathbb{R}^2$ and $\psi : \mathbb{R}^2 \to \mathbb{R}$ is a known smooth function modeling ground topography. Without loss of generality, we let \textbf{x} be the position of the targets at the beginning of the synthetic aperture, at time $s_0$ = 0. Assuming the scatterer moves at a constant velocity over the acquisition process, we can represent the trajectory of the scatterer as 
\begin{equation}	
\mathbf{z}(s) = \mathbf{x} + \mathbf{v_x}s,
\end{equation}
where $\mathbf{v_x} \in \mathbb{R}^3$ is the velocity of a particular point scatterer located at point \textbf{x} at time $s_0$ = 0. Since the target is moving on the surface, the velocity $\mathbf{v_x}$ is of the form
\begin{equation}
 \mathbf{v_x} = [\pmb{\nu_\textit{\textbf{x}}}, \nabla_\textit{\textbf{x}}\psi(\textit{\textbf{x}})\cdot\pmb{\nu_\textit{\textbf{x}}}],
\end{equation}
where $\pmb{\nu_\textit{\textbf{x}}}$ is the 2D-velocity of the target and $\nabla_\textit{\textbf{x}}\psi(\textit{\textbf{x}})$ is the gradient of the ground topography. 

We define the phase-space reflectivity function of a target as: 
\begin{align}
q(\textit{\textbf{x}},\pmb{\nu}) &= \rho (\textit{\textbf{x}})\delta (\pmb{\nu} - \pmb{\nu_x})\\
&\approx \rho (\textit{\textbf{x}}) \varphi (\pmb{\nu},\pmb{\nu_x})
\end{align}
where $\rho(\textit{\textbf{x}})$ represents the 2D surface reflectivity of a point on the ground, and $\varphi (\pmb{\nu},\pmb{\nu_x})$ is a smooth, differentiable function of $\pmb{\nu}$ that approximates the Dirac
delta function in the limit. Under Born approximation, the received signal model for monostatic SAR is defined as follows \cite{duman2015moving,wang2012bistatic,wang2013ground}:
\begin{multline}\label{eq:received_mono}
d(s,t) = \int \text{e}^{-\text{i}\phi(\omega,s,t,\pmb{x},\pmb{\nu})} A(\omega,s,\textit{\textbf{x}},\pmb{\nu})q(\textit{\textbf{x}},\pmb{\nu}) d\omega d\textit{\textbf{x}} d\pmb{\nu}
\end{multline}
where $t$ is the fast-time, $\omega \in \mathbb{R}$ is the temporal frequency, and $A(\omega,s,\textit{\textbf{x}},\pmb{\nu})$ is a complex amplitude function varying slowly in $\omega$. It includes transmitter and receiver antenna beam patterns, the transmitted waveforms, and geometrical spreading factors. $\phi(\omega,s,t,\textit{\textbf{x}},\pmb{\nu})$ is the phase function given by
\begin{equation}
\phi(\omega,s,t,\pmb{x},\pmb{\nu}) = \frac{\omega}{c_0}(t-(R(s,\pmb{x})+B(s,\pmb{x},\pmb{\nu}))),
\end{equation}  with $R(s,\textit{\textbf{x}})$ being the range term, $B(s,\textit{\textbf{x}},\pmb{\nu})$ is the range variation due to the movement of the scatterer $\textit{\textbf{x}}$, and $c_0$ the speed of light. For a monostatic SAR configuration, 
\begin{equation}
R(s,\pmb{x}) = 2|\gamma(s) - \textbf{x}|,
\end{equation}
\begin{equation}
B(s,\pmb{x},\pmb{\nu}) = 2s[(\widehat{\textbf{x}-\gamma(s)})\cdot \mathbf{v}],
\end{equation} where $\gamma(s) \in \mathbb{R}^3$ is the location of the transmitter/receiver pair, and $(\widehat{\textbf{x}-\gamma(s)})$ denotes the unit vector from the transmitter/receiver to the scatterer.

\subsection{Discretized Model}
We discretize the underlying scene reflectivity function and represent it in the standard pixel basis such that the image domain is sampled on an $N$-dimensional grid of $D := \{\textbf{\textit{x}}_k\}_{k=1}^N$. Furthermore, we discretize the range of possible velocities using $[\Delta \nu_1, \Delta \nu_2]$ intervals into $M$-samples to obtain an $M \times N$ sized matrix $\mathbf{Q}$:
\begin{equation}\label{eq:PSR}
\mathbf{Q} = \begin{bmatrix}
\rho(\textit{\textbf{x}}_1)\delta(\pmb{\nu}_{1} - \pmb{\nu}_{\textit{\textbf{x}}_1}) 
&  \cdots \ \rho(\textit{\textbf{x}}_N)\delta(\pmb{\nu}_{1} - \pmb{\nu}_{\textit{\textbf{x}}_N})\\

\rho(\textit{\textbf{x}}_1)\delta(\pmb{\nu}_{2} - \pmb{\nu}_{\textit{\textbf{x}}_1}) 
&  \cdots \ \rho(\textit{\textbf{x}}_N)\delta(\pmb{\nu}_{2} - \pmb{\nu}_{\textit{\textbf{x}}_N}) \\

\vdots & \vdots \qquad \qquad \qquad \qquad \vdots\\
\rho(\textit{\textbf{x}}_1)\delta(\pmb{\nu}_{M} - \pmb{\nu}_{\textit{\textbf{x}}_1}) 
& \cdots \ \rho(\textit{\textbf{x}}_N)\delta(\pmb{\nu}_{M} - \pmb{\nu}_{\textit{\textbf{x}}_N})\\
\end{bmatrix}.
\end{equation}

We refer to $\mathbf{Q}$ in \eqref{eq:PSR} as the PSR matrix. Since each target can only be moving at a single velocity, each column contains only one non-negative entry. Each row can contain multiple non-negative entries, as multiple targets may move at the same velocities. Since all stationary targets have the same velocity $\pmb{\nu} = 0 $, they appear in the same \textit{row} of $\mathbf{Q}$ corresponding to some where index $\pmb{\nu}_i = 0$. We denote this index as $\nu_s$. Thus, stationary targets form a rank-one component in $\mathbf{Q}$, which we denote as $\mathbf{Q}_s$. 

Separating $\mathbf{Q}_s$ from $\mathbf{Q}$ leaves the terms that correspond to moving targets, which we denote as $\mathbf{Q}_{\nu}$. The columns of $\mathbf{Q}_{\nu}$ span a subspace in $\mathbb{C}^{M}$ of dimension equal to the number of unique velocities within the scene of interest, which is upper bounded by the number of moving targets. 
As a result, the PSR matrix $\mathbf{Q}$ consists of a rank-one $\mathbf{Q}_s$ and sparse $\mathbf{Q}_\nu$ as follows:
\begin{equation}
\mathbf{Q} = \mathbf{Q}_s + \mathbf{Q}_\nu, \quad \text{where} \ \mathbf{Q}_s = \mathbf{e}_{\nu_s} \pmb{\rho}_s^H.
\end{equation}
The $\mathbf{Q}_\nu$ term includes the reflectivities and velocities of the moving targets, as well as their locations at time $s=0$. The $\mathbf{e}_{\nu_s}$ term is the standard column basis vector in $M$ that has the $1$ entry at the index corresponding to $\nu_s$, $0$ else, and $\pmb{\rho}_s$ is the vector of unknown stationary reflectivity coefficients. 

It is important to note that given that there can only be a single non-zero element per column of $\mathbf{Q}_\nu$, the sparsity assumption holds even if moving targets are dense on the imaging grid. Meanwhile, a moving target is considered stationary if it does not move out of a single range resolution cell by the end of the aperture. That is, given that the trajectory of a scatterer is modeled by \eqref{eq:traj}, the velocity of any given target $\pmb{x}_k$ must satisfy
\begin{equation}\label{eq:traj}
|\pmb{\nu}_{\pmb{x}_k}| > \frac{c_0}{2BS},
\end{equation}
where $\frac{c_0}{2B}$ is the range resoltuion, and $S$ is the total data acquisition time over the aperture.

To illustrate the structure in $\mathbf{Q}$, we consider the following example. Let \textit{\textbf{x}}$_1$, \textit{\textbf{x}}$_2\ldots$, \textit{\textbf{x}}$_N$ be locations of targets on the scene, and let the range of 2D velocities [$\pmb{\nu}_{min}, \pmb{\nu}_{max}$] be discretized into $\pmb{\nu}_1 = \pmb{\nu}_{min},\ \pmb{\nu}_2,\ \pmb{\nu}_3 = 0,\ \pmb{\nu}_4,\ \pmb{\nu}_5 = \pmb{\nu}_{max}$. In this case, $M = 5$. Let \textit{\textbf{x}}$_2$, \textit{\textbf{x}}$_k$ be the locations of moving targets at time $s=0$, moving linearly with velocities $\pmb{\nu}_4,\ \text{and}\ \pmb{\nu}_2$, respectively, and let the remaining targets be stationary. Then, the phase-space reflectivity matrices take on the form:
\begin{equation}
\begin{split}
\quad &\mathbf{Q}_s = 
\setlength\arraycolsep{2pt}
\begin{bmatrix}
0 & 0 & 0 & \ldots & 0 & 0 & 0 & \dots & 0\\
0 & 0 & 0 & \ldots & 0 & 0 & 0 & \dots & 0\\
\rho(\textit{\textbf{x}}_1) & 0 & \rho(\textit{\textbf{x}}_3) & \ldots & \rho(\textit{\textbf{x}}_{k-1}) & 0 & \rho(\textit{\textbf{x}}_{k+1}) & \dots & \rho(\textit{\textbf{x}}_N)  \\
0 & 0 & 0 & \ldots & 0 & 0 & 0 & \dots & 0\\
0 & 0 & 0 & \ldots & 0 & 0 & 0 & \dots & 0
\end{bmatrix}\\
\\
\quad &\mathbf{Q}_\nu = 
\setlength\arraycolsep{5pt}
\begin{bmatrix}
0 & 0 & 0 & \ldots & 0 & 0 & 0 &\ldots & 0\\
0 & 0 & 0 & \ldots & 0 & \rho(\textit{\textbf{x}}_k) & 0 &\ldots & 0\\
0 & 0 & 0 & \ldots & 0 & 0 & 0 &\ldots & 0\\
0 & \rho(\textit{\textbf{x}}_2) & 0 & \ldots & 0 & 0 & 0 &\ldots & 0\\
0 & 0 & 0 & \ldots & 0 & 0 & 0 &\ldots & 0
\end{bmatrix}.
\end{split}
\end{equation}

Next, we discretize the received signal model and define the kernel as
\begin{equation}\label{eq:kernel}
\begin{split}
F(\omega_l, s_m, t_n,\textit{\textbf{x}}_k, \pmb{\nu}_{k'}) = \text{e}^{-\text{i}\phi(\omega_l, s_m, t_n, \textit{\textbf{x}}_k,\pmb{\nu}_{k'})}\times\\  A(\omega_l,s_m,\textit{\textbf{x}}_k,\pmb{\nu}_{k'}),
\end{split}
\end{equation}
The resulting discretized model is expressed as
\begin{equation}\label{eq:discrete_model}
\textit{d}(s_m,t_n) \approx \sum_{l=1}^L\sum_{k=1}^N \sum_{k' = 1}^{M} F(\omega_l, s_m,t_n, \textit{\textbf{x}}_k,\pmb{\nu}_{k'})\mathbf{Q}(\textit{\textbf{x}}_k,\pmb{\nu}_{k'}).
\end{equation}
Letting \textbf{F}$(s_m,t_n)$ denote the discrete $M \times N$ sized matrix whose kernel is defined in (\ref{eq:discrete_model}) for a fixed $ s_m,t_n$, the SAR data in the presence of moving targets corresponds to a linear model that is measured via a Frobenius inner-product with the underlying unknown matrix $\mathbf{Q}$ as:
\begin{equation} \label{eq:forward_map}
\textit{d}(s_m,t_n) := \langle \mathbf{F}( s_m,t_n), \mathbf{Q} \rangle_F,
\end{equation}
where $\mathbf{Q}$ is the sum of a rank-one $\mathbf{Q}_s$ and sparse $\mathbf{Q}_v$. The linear mapping from the unknown matrix $\mathbf{Q}$ to the collection of SAR measurements $\{ \textit{d}(s_m,t_n) \}$ sampled along the flight trajectory of the receiver can be expressed via a \emph{lifted forward model}, $\mathcal{F}$, such that
\begin{equation}\label{eq:lifted_forward_model}
\mathbf{d} :=  \mathcal{F}(\mathbf{Q}),
\end{equation} 
where $\mathbf{d}$ is formed by stacking the SAR measurements $\textit{d}(s_m,t_n)$ into a vector. 

In the following sections, we derive algorithms used to detect moving targets and estimate their velocities.

\section{Moving Target Imaging and Velocity Estimation} \label{algorithms}
In this section, we construct our objective function and solve the resulting optimization problem via proximal gradient descent and alternating direction method of multipliers to simultaneously recover stationary and moving targets in $\mathbf{Q}_s$ and $\mathbf{Q}_\nu$, respectively. We define the constraint $C_1$ as the set of rank-one matrices with a known right singular vector $\mathbf{e}_{\nu_s}$. Recall that the discretization of velocities is known, and thereby both $\nu_s$ and the support of $C_1$ are known. Therefore,
\begin{equation}
C_1 = \{\mathbf{e}_{\nu_s}\mathbf{x}^H \in \mathbb{R}^{M\times N}_+ : \mathbf{x} \in \mathbb{R}^N\}.
\end{equation}
We define the indicator function $\mathcal{I}_{C_1}(\mathbf{X})$ as
\begin{equation}
\mathcal{I}_{C_1}(\mathbf{X}) = \begin{cases}
0 \quad \text{if}\ \mathbf{X}\in C_1\\
\infty \quad \text{if}\ \mathbf{X} \notin C_1.
\end{cases}
\end{equation}
We introduce a second convex constraint set as the set of matrices that lie in the orthogonal complement of $C_1$, denoted as $C_1^c$, and define the indicator function accordingly. Therefore, we consider the sparse recovery problem:
\begin{align} \label{eq:final_obj}
\min_{\mathbf{Q}_s,\mathbf{Q}_\nu \in \mathbb{R}^+} \quad & \|\mathbf{Q}_\nu\|_1\\
\textrm{s.t} \quad & \|\textbf{d} - \mathcal{F}(\mathbf{Q}_s +\mathbf{Q}_\nu)\|_F \leq \delta \nonumber\\
\quad & \mathbf{Q}_s \in C_1,\ \mathbf{Q}_\nu \in C_1^c, \nonumber
\end{align}
for some noise level $\delta >0 $.
%\textcolor{red}
{Ultimately, the objective is to recover a sparse matrix $\mathbf{Q}_\nu \in \mathbb{R}^{M\times N}$ that consists of the reflectivities of the moving targets in their correct position at time $s=0$, while abiding by the constraints that incorporate the stationary target estimates. By searching over the rows of $\mathbf{Q}_\nu$, we also obtain the 2D velocity estimates of each target.} Furthermore, we also recover the rank-one stationary component $\mathbf{Q}_s \in \mathbb{R}^{M \times N}$ that consists of the entire stationary scene. To do so, we derive iterative solutions via PGD and ADMM.

Note that while the structure in $\mathbf{Q}_s$ and $\mathbf{Q}_\nu$ are similar to the problem setup of RPCA, we have a particularly favorable structure in the PSR matrix which simplifies the problem and reduces computational complexity. For our problem formulation over $\mathbf{Q}$, solving the generic RPCA amounts to a complexity of $\text{min} (\mathcal{O}(MN^2),\mathcal{O}(M^2N))$ due to the singular-value threshold step, presenting a bottleneck when considering high-resolution images that are typically used in SAR imaging. 
\subsection{PGD for PSR Matrix Recovery}
Forming the unconstrained penalty function for \eqref{eq:final_obj}, we obtain:
\begin{equation}\label{eq:pgd_obj}
\begin{split}
\mathcal{L}(\mathbf{Q}_s,\mathbf{Q}_\nu) = \lambda\|\mathbf{Q}_\nu\|_1 + \frac{r}{2}\|\mathcal{F}(\mathbf{Q}_s + \mathbf{Q}_\nu) - \textbf{d}\|_2^2\\ + \mathcal{I}_{C_1}(\mathbf{Q}_s) + \mathcal{I}_{C_1^c}(\mathbf{Q}_\nu),
\end{split}
\end{equation}
where $r>0$ is a penalty parameter related to the noise level $\delta$, and $\lambda$ is a regularization parameter that controls the sparsity of $\mathbf{Q}_\nu$. This can be solved by iterative shrinkage thresholding algorithm (ISTA) \cite{daubechies2004iterative,daubechies2007iteratively} in parallel over $\mathbf{Q}_s$ and $\mathbf{Q}_\nu$ variables by stacking them, and applying the respective proximal maps block-wise after the descent step over the quadratic term. 
That is, given $\underset{\mathbf{X}}{\text{min}}\ F(\mathbf{X}) = f(\mathbf{X}) + g(\mathbf{X})$, where $f(\mathbf{X})$ is $L-$smooth and $g(\mathbf{X})$ is closed and convex
\begin{equation}
\mathbf{X}^{k+1} = \text{prox}_{\alpha_k g}(\mathbf{X}^k - \alpha_k \nabla(f(\mathbf{X}^k)),
\end{equation}
where $\alpha_k$ is a step-size bounded by the inverse of the Lipschitz constant of $f(\mathbf{X})$. As long as $\alpha_k$ is bounded in this fashion, PGD is guaranteed to converge at a linear rate of $\mathcal{O}(1/\epsilon)$ \cite{Ma2018}. Note that in the block-wise form, the support constraint of $\mathbf{Q}_s$, merely corresponds to a block over which only the smooth gradient step is performed. This is identical to a projection in the matrix notation, denoted by $\mathcal{P}_{C_1}(\cdot)$ in Algorithm \ref{alg:PGD}. 
Meanwhile, the $\mathbf{Q}_\nu$ update falls under a class of proximal functions which are summable \cite{yu2013decomposing}, that is 
\begin{equation}
\text{prox}_{g+h} = \text{prox}_h \circ \text{prox}_g.
\end{equation}

Letting $g(\mathbf{X}) = \lambda \|\mathbf{X}\|_1$ and $h(\mathbf{X}) = \mathcal{I}_{C_1^c}(\mathbf{X})$, the update term becomes
\begin{equation}
\mathbf{X}^{k+1} = \mathcal{P}_{C_1^c} (\text{ST}_\frac{\lambda}{\alpha_k}(\mathbf{X}^k - \alpha_k \nabla f(\mathbf{X}^k))),
\end{equation}  
where $\mathcal{P}_{C_1^c}(\cdot)$ is the projection onto the set $C_1^c$, and $\text{ST}(\mathbf{X})$ is the soft-threshold operator. Observe that both projections can be efficiently implemented via selection operators. Specifically, $\mathcal{P}_{C_1}(\cdot)$ is implemented by selecting the row in $\mathbf{Q}$ that corresponds to stationary scatterers, and $\mathcal{P}_{C_1^c}(\cdot)$ selects the remaining rows. The final algorithm for PGD can be found in Algorithm \ref{alg:PGD}. By replacing the nuclear-norm constraint on $\mathbf{Q}_s$ with our rank-one constraint, we see the update step no longer requires a singular value decomposition.
\\
\begin{algorithm}
\caption{Proximal Gradient Descent}
\label{alg:PGD}
\begin{algorithmic}

\STATE{\textbf{Step 0.}} Initialize $\mathbf{Q}_s^0,\mathbf{Q}_\nu^0,r,\alpha_k, \delta$ 
 \WHILE{$\|\mathbf{d}-\mathcal{F}(\mathbf{Q}^k)\| >\ \delta$}
\STATE{\textbf{Step 1.}} $\begin{aligned}[t]&\nabla f_{\mathbf{Q}_s}(\mathbf{Q}_s^k,\mathbf{Q}_\nu^k) = r \mathcal{F}^H(\mathcal{F}(\mathbf{Q_s}^k + \mathbf{Q}_\nu^k) - \mathbf{d})\\
&\nabla f_{\mathbf{Q}_\nu}(\mathbf{Q}_s^k,\mathbf{Q}_\nu^k) = r \mathcal{F}^H(\mathcal{F}(\mathbf{Q_s}^k + \mathbf{Q}_\nu^k) - \mathbf{d})
\end{aligned}$
\STATE{\textbf{Step 2.}} $\mathbf{Q}_s^{k+1} = \mathcal{P}_{C_1}(\mathbf{Q}_s^k - \alpha^k\nabla f_{\mathbf{Q}_s}(\mathbf{Q}_s^k,\mathbf{Q}_\nu^k))$
\STATE{\textbf{Step 3.}} $\mathbf{Q}_\nu^{k+1} = \mathcal{P}_{C_1^c}(\text{ST}_{\frac{\lambda}{\alpha_k}}(\mathbf{Q}_\nu^k - \alpha^k \nabla f_{\mathbf{Q}_\nu}(\mathbf{Q}_s^k,\mathbf{Q}_\nu^k))$
\STATE{\textbf{Step 4.}} Update $\alpha^k$ through backtracking line search
\STATE{\textbf{Step 5.}} $k = k+1$

\ENDWHILE
\end{algorithmic}
\end{algorithm}
	
While PGD is well-studied and has been proven to converge using simple update equations, its convergence can be quite slow. Furthermore, the step-size $\alpha_k$ can be difficult to determine. We recognize the problem formulation in \eqref{eq:final_obj} can be adapted as a standard sparse recovery problem over $\mathbf{Q}$, where the soft-threshold operator would be performed in a suitable block-wise manner. While the two problems are equivalent, the current formulation proves to be more informative of the underlying mathematical structure of the PSR matrix. To improve the convergence rate, FISTA-based \cite{beck2009fast,chambolle2015convergence} updates can be deployed. Additionally, convergence speed up and parameter tuning can be addressed through LISTA \cite{gregor2010learning,monga2021algorithm} providing an enticing direction using the PGD solvers.
\subsection{ADMM for PSR Matrix Recovery}
Alternating direction method of multipliers is a variation of the augmented Lagrangian method. ADMM is particularly favorable when the objective function is separable, allowing for an alternating scheme as opposed to the typical joint minimization that methods such as PGD deploy \cite{boyd2011distributed}. Hence, ADMM has been a typical choice for generic RPCA problems. In addition, ADMM has shown improved empirical performance, demonstrating convergence in much fewer iterations than PGD although theory suggests identical rates for the two methods \cite{Ma2018}. 
Here, we explore ADMM as a viable option for our formulation, which admits a suitable block-wise form for the solver.
 
To simplify the algorithmic procedure of ADMM, we introduce the dummy variable $\mathbf{Q}$, and formulate the problem as:
\begin{equation}\label{eq:reformulated_opt}
\begin{aligned}
		\min \quad & \lambda \| \mathbf{Q}_\nu \|_1 +\frac{1}{2}\|\mathcal{F}(\mathbf{Q}) - \mathbf{d} \|_2^2 +  \mathcal{I}_{C_1}(\mathbf{Q}_s) + \mathcal{I}_{C_1^c}(\mathbf{Q}_\nu)\\ 
		\textrm{s.t} \quad & \mathbf{Q} = \mathbf{Q}_s + \mathbf{Q}_\nu.
	\end{aligned}
\end{equation}
This formulation is equivalent to \eqref{eq:final_obj} in its exact solution, but facilitates the splitting of the following augmented Lagrangian:
\begin{equation}\label{eq:new_lagrang}
\begin{split}
\hat{\mathcal{L}}_r(\mathbf{Q},\mathbf{Q}_s,\mathbf{Q}_\nu,\textbf{Y}) =\lambda \|\mathbf{Q}_\nu\|_1 + \frac{1}{2}\|\mathcal{F}(\mathbf{Q}) - \textbf{d}\|_2^2\\ + \langle \textbf{Y}, \mathbf{Q} - \mathbf{Q}_s - \mathbf{Q}_\nu \rangle_F\\ + \frac{r}{2}\|\mathbf{Q} - \mathbf{Q}_s- \mathbf{Q}_\nu \|^2_F + \mathcal{I}_{C_1}(\mathbf{Q}_s) + \mathcal{I}_{C_1^c}(\mathbf{Q}_\nu),
\end{split}
\end{equation}
where $\mathbf{Y}$ are the Lagrange multipliers, and we have assumed $\mathbf{Q},\mathbf{Q}_s,\mathbf{Q}_\nu$,and $\mathbf{Y}$ to be constrained as real-valued. It is important to note that two-block ADMM for convex problems globally converges for any penalty parameter $r > 0$ \cite{eckstein1992douglas}. Using the variable splitting technique, an approximate closed-form solution for $\mathbf{Q}$ can be obtained via
\begin{equation}
\begin{split}
\mathbf{Q}^{k+1} &= (\mathcal{F}^H\mathcal{F}+r\mathbf{I})^{-1}(\mathcal{F}^H\textbf{d}-\textbf{Y}^k + r(\mathbf{Q}_s^k + \mathbf{Q}_\nu^k)),\\
&\approx \frac{1}{1+r}(\mathcal{F}^H\textbf{d}-\textbf{Y}^k + r(\mathbf{Q}_s^k + \mathbf{Q}_\nu^k))
\end{split}
\end{equation}
where we use the approximation $\mathcal{F}^H\mathcal{F} \approx \mathbf{I}$ \cite{wacks2014passive_part2,duman2015moving} to avoid the demanding inversion step. Using the updated $\mathbf{Q}^{k+1}$, the ADMM iterations then reduce to solving for the unknowns $\mathbf{Q}_s$ and $\mathbf{Q}_\nu$, such that:
\begin{equation}
\mathbf{Q}_s^{k+1} := \underset{\mathbf{Q}_s}{\text{argmin}} \ \hat{\mathcal{L}}_r(\mathbf{Q}^{k+1},\mathbf{Q}_s,\mathbf{Q}_\nu^k,\textbf{Y}^k)
\end{equation}
\begin{equation}\label{eqref:ADMM_Qs}
\mathbf{Q}_\nu^{k+1} := \underset{\mathbf{Q_\nu}}{\text{argmin}}\ \hat{\mathcal{L}}_r(\mathbf{Q}^{k+1},\mathbf{Q}_s^{k+1},\mathbf{Q}_\nu,\textbf{Y}^k)
\end{equation}
\begin{equation}
\textbf{Y}^{k+1} := \underset{\mathbf{Y}}{\text{argmin}} \ \hat{\mathcal{L}}_r(\mathbf{Q}^{k+1},\mathbf{Q}_s^{k+1},\mathbf{Q}_\nu^{k+1},\textbf{Y})
\end{equation}
in which the update steps for $\mathbf{Q}_s$ and $\mathbf{Q}_\nu$ feature the respective proximity operators. The ADMM update steps can be found in Algorithm \ref{alg:updated_admm_rpca}.
\begin{algorithm}
\caption{Alternating Direction Method of Multipliers}
\label{alg:updated_admm_rpca}
\begin{algorithmic}

\STATE{\textbf{Step 0.}} Initialize $\mathbf{Q}^0,\mathbf{Q}_s^0,\mathbf{Q}_\nu^0,\textbf{Y}^0,r, \delta$ 
 \WHILE{$\|\textbf{d}-\mathcal{F}(\mathbf{Q}^k)\| >\ \delta$}
\STATE{\textbf{Step 1.}} $\mathbf{Q}^{k+1} = \frac{1}{1+r}(\mathcal{F}^H\textbf{d}-\textbf{Y}^k + r(\mathbf{Q}_s^k + \mathbf{Q}_\nu^k))$
\STATE{\textbf{Step 2.}} $\mathbf{Q}_s^{k+1} = \mathcal{P}_{C_1}(\mathbf{Q}^{k+1}-\mathbf{Q}_\nu^{k}+\textbf{Y}^{k})$
\STATE{\textbf{Step 3.}} $\mathbf{Q}_\nu^{k+1} = \mathcal{P}_{C_1^c}(ST_{\frac{\lambda}{r}}(\mathbf{Q}^{k+1}-\mathbf{Q}_s^{k+1}+\textbf{Y}^{k}))$
\STATE{\textbf{Step 4.}} $\textbf{Y}^{k+1} = \textbf{Y}^{k} + \frac{1}{r}(\mathbf{Q}^{k+1}-\mathbf{Q}_s^{k+1}-\mathbf{Q}_\nu^{k+1})$
\STATE{\textbf{Step 5.}} $k = k+1$

\ENDWHILE
\end{algorithmic}
\end{algorithm}
\subsection{Prior Knowledge of Number of Moving Targets}
{Here, we discuss a non-convex approach to GMTI. When the number of moving targets is known a priori, the number of false alarms can be reduced by imposing a hard cardinality constraint on $\mathbf{Q}_\nu$, that is card($\mathbf{Q}_\nu) \leq k$. This can be achieved by replacing the $\ell_1-$ norm with the $\ell_0-$norm and solving   
\begin{align} \label{eq:non_convex}
\min_{\mathbf{Q}_s,\mathbf{Q}_\nu \in \mathbb{R}^+} \quad &\|\mathbf{Q}_\nu\|_0\\\
\textrm{s.t} \quad & \|\textbf{d} - \mathcal{F}(\mathbf{Q}_s +\mathbf{Q}_\nu)\|_F, \nonumber \\& \mathbf{Q}_s \in C_1,\ \mathbf{Q}_\nu \in C_1^c,\nonumber\\& \text{card}(\mathbf{Q_\nu}) \leq k. \nonumber
\end{align}
Again, noting that $\mathbf{Q}_s$ and $\mathbf{Q}_\nu$ are easily separable, \eqref{eq:non_convex} can be solved via
\begin{align} \label{eq:non-convex Qs}
\mathbf{Q}_s^{k+1} &= \underset{\mathbf{Q}_s}{\text{argmin}}\ \frac{r}{2}\|\mathcal{F}(\mathbf{Q}_s + \mathbf{Q}_\nu) - \textbf{d}\|_2^2  + \mathcal{I}_{C_1}(\mathbf{Q}_s)
\end{align}
\begin{equation}
\begin{split}
\mathbf{Q}_\nu^{k+1} = \underset{\text{card}(\mathbf{Q}_\nu) \leq k}{\text{argmin}}\ \|\mathbf{Q}_\nu\|_0 &+ \frac{r}{2}\|\mathcal{F}(\mathbf{Q}_s + \mathbf{Q}_\nu) - \textbf{d}\|_2^2 \\
&\quad + \mathcal{I}_{C_1^c}(\mathbf{Q}_\nu)
\end{split}\label{eq:non-convex Qnu}
\end{equation}
Here, the subproblem in \eqref{eq:non-convex Qs} is solved in exactly the same manner as the convex case, and \eqref{eq:non-convex Qnu} can be solved by applying a hard-threshold to the gradient descent step. The proximal operator corresponds to a projection onto the set of $k$-largest elements, $\Omega$, which directly enforces the cardinality constraint. We denote this projection as $\mathcal{P}_\Omega (\cdot)$. The update steps are found in Algorithm \ref{alg:non-convex alg}.}
\begin{algorithm}
\caption{Non-Convex Proximal Gradient Descent}
\label{alg:non-convex alg}
\begin{algorithmic}

\STATE{\textbf{Step 0.}} Initialize $\mathbf{Q}_s^0,\mathbf{Q}_\nu^0,r,\alpha_k, \delta$ 
 \WHILE{$\|\mathbf{d}-\mathcal{F}(\mathbf{Q}^k)\| >\ \delta$}
\STATE{\textbf{Step 1.}} $\begin{aligned}[t]&\nabla f_{\mathbf{Q}_s}(\mathbf{Q}_s^k,\mathbf{Q}_\nu^k) = r \mathcal{F}^H(\mathcal{F}(\mathbf{Q_s}^k + \mathbf{Q}_\nu^k) - \mathbf{d})\\
&\nabla f_{\mathbf{Q}_\nu}(\mathbf{Q}_s^k,\mathbf{Q}_\nu^k) = r \mathcal{F}^H(\mathcal{F}(\mathbf{Q_s}^k + \mathbf{Q}_\nu^k) - \mathbf{d})
\end{aligned}$
\STATE{\textbf{Step 2.}} $\mathbf{Q}_s^{k+1} = \mathcal{P}_{C_1}(\mathbf{Q}_s^k - \alpha^k\nabla f_{\mathbf{Q}_s}(\mathbf{Q}_s^k,\mathbf{Q}_\nu^k))$
\STATE{\textbf{Step 3.}} $\mathbf{Q}_\nu^{k+1} = \mathcal{P}_{C_1^c}(\mathcal{P}_\Omega(\mathbf{Q}_\nu^k - \alpha^k \nabla f_{\mathbf{Q}_\nu}(\mathbf{Q}_s^k,\mathbf{Q}_\nu^k))$
\STATE{\textbf{Step 4.}} Update $\alpha^k$ through backtracking line search
\STATE{\textbf{Step 5.}} $k = k+1$

\ENDWHILE
\end{algorithmic}
\end{algorithm}
\subsection{Computational Complexity}
{We now present the computational complexity of the image reconstruction algorithm. We assume that there are $\mathcal{O}(N)$ samples in both slow-time and fast-time, and the scene of interest is sampled in $\mathcal{O}(N)$. We also assume there are $M$ different hypothesized velocities used for reconstruction. }
{We begin with a computational complexity analysis for PGD:
\begin{itemize}
\item{\textit{Calculation of the gradients}: This step is dominated by the calculation of $\mathcal{F}^H\mathbf{d}$, which requires $\mathcal{O}(MN^3)$ operations. Note that $\mathcal{F}$ becomes a Fourier Integral Operator (FIO) \cite{duman2015moving,trevesintroduction,hormander1971fourier} and can be implemented using fast back-projection algorithms reducing complexity to $\mathcal{O}(MN^2 \log N)$ \cite{candes2007fast,candes2009fast}. Meanwhile, this quantity can be calculated and stored without the need of recalculation. In our implementations, we make the reasonable approximation $\mathcal{F}^H\mathcal{F}$ is identity \cite{wacks2014passive_part2,duman2015moving}. Thus, the computational complexity for this step is that of matrix addition, $\mathcal{O}(MN)$}
\item{\textit{$\mathbf{Q}_s$ update step:} In comparison to generic RPCA, this step is analogous to updating the low-rank component. However we no longer require a singular value decomposition. Note that the projection $\mathcal{P}_{C_1}(\cdot)$ simply retains the \emph{row} support, corresponding to the index where $\nu_s = 0$. This requires $\mathcal{O}(N)$ operations. Thus, this step takes $\mathcal{O}(MN)$ operations.}
\item{\textit{$\mathbf{Q}_\nu$ update step:} The update step for $\mathbf{Q}_\nu$ requires a soft-thresholding operator, which acts as a point-wise operator. Meanwhile, the projection $\mathcal{P}_{C_1^c}(\cdot)$ can be implemented by zeroing out the row corresponding to the index where $\nu_s = 0$, which requires $\mathcal{O}(N)$ operations. Thus, this step requires $\mathcal{O}(MN)$ operations.}
\end{itemize}
Therefore, the per-iteration cost of performing PGD is $\mathcal{O}(MN)$, an order of magnitude lower than generic RPCA. 
{We next discuss the computational complexity analysis for ADMM.
\begin{itemize}
\item{\textit{$\mathbf{Q}$ update step:} The update step is again dominated by the calculation of $\mathcal{F}^H\mathbf{d}$. However, like PGD, this only needs to be calculated once and can be stored. Thus, the complexity of this step is that of matrix addition, $\mathcal{O}(MN)$.}
\item{\textit{$\mathbf{Q}_s$ and $\mathbf{Q}_\nu$ update steps:} The proximal operators used in ADMM are the same as PGD. Meanwhile, the arguments of the operators remain as matrix addition. Thus, this step has a complexity of $\mathcal{O}(MN)$.}
\item{\textit{$\mathbf{Y}$ update step:} The update step for $\mathbf{Y}$ only requires matrix addition, leading to a complexity of $\mathcal{O}(MN)$}.
\end{itemize}
Thus, ADMM has a per-iteration complexity of $\mathcal{O}(MN)$. Finally, we discuss the computational complexity of the non-convex approach. The update steps are the same as the steps outlined in Algorithm \ref{alg:PGD}, with the exception of the hard-thresholding step in the non-convex case. Thus, the complexity of the non-convex approach is as follows:
\begin{itemize}
\item{\textit{Calculation of the gradients}: As stated previously, this step requires $\mathcal{O}(MN)$ operations per-iteration.}
\item{\textit{$\mathbf{Q}_s$ update step:} As stated previously, this step requires $\mathcal{O}(N)$ operations per-iteration.}
\item{\textit{$\mathbf{Q}_\nu$ update step:} This step replaces the soft-thresholding operator with the cardinality constraint. In order to obtain the $k-$largest values of $\mathbf{Q}_\nu$, the matrix must first be sorted. Thus, this step requires $\mathcal{O}(MN\log MN)$ operations per-iteration.}
\end{itemize}
From this, we see that incorporating a priori knowledge on the cardinality of $\mathbf{Q}_\nu$ to reduce the number of false alarms at the cost of non-convexity and a higher computational cost of $\mathcal{O}(MN\log MN)$. This is still, however, less than that of generic RPCA.} Meanwhile, we see that both PGD and ADMM have the same order of complexity, up to a constant scaling factor.

\section{Numerical Experiments} \label{Numerical Experiments}
We performed four sets of numerical simulations to demonstrate the performance of our imaging method. In the first set of simulations, we studied the performance of reconstruction and velocity estimation for  multiple moving point targets when the received data is contaminated with additive noise. In the second set of simulations, we study the robustness of the algorithms in reconstruction and velocity estimation when the moving targets are embedded in clutter, varying the signal-to-clutter ratio. In the third set of simulations, we study the robustness of the algorithms when both the received data is contaminated by additive noise and the moving targets are embedded in clutter by varying the signal-to-clutter-plus-noise ratio (SCNR). Finally, in the fourth set of simulations, we studied the robustness of the algorithm with respect to the number of moving targets in the scene. 
\subsection{Scene and Imaging Parameters}
For all numerical simulations, we consider a monostatic antenna traversing a circular trajectory. We considered a scene of size 100 $\times$ 100m$^2$ scene with flat topography. The scene was discretized into 31 $\times$ 31 pixels such that each pixel represents approximately 3.25m, where [0,0] m and [100,100] m correspond to the pixels (1,1) and (31,31), respectively. The antenna traverses a circular trajectory defined as $\gamma(s) = [11 + 11\cos (s),11 + 11\sin (s), 6.5]$km at a constant speed, where $\nu$ = 950 km/hr, completing a full aperture in approximately 262.5 s. The center frequency is 9 GHz with a bandwidth of 50 MHz, such that the ground-range resolution is approximately 3m. The slow-time and fast-time are sampled uniformly, obtaining 512 and 100 samples respectively. We summarize key imaging and scene parameters for the simulation set-ups that other methods have utilized in Table \ref{tab:parameters}. Note that \cite{yang2015strong,li2016moving} operate on real-data, and thus the number of moving targets nor the clutter type is unknown. 

For each simulation unless otherwise stated, we assumed that there were twelve moving point targets moving in the range of [-20,20] $\times$ [-20,20] m/s, within the range of velocities that is expected from ground moving targets such as vehicles. We discretized the velocity space into a 21 $\times$ 21 grid with a step size of 2 m/s, resulting in 441 unique velocities. In order to demonstrate robustness with respect to spatial resolution and velocity resolution, we make note of the following scene parameters:
\begin{itemize}
\item{We assumed two targets located next to each other in pixels [10,5] and [10,6] at time $s=0$, moving at velocities [14,12] and [2,-4] m/s respectively.}
\item{We assumed two targets located far apart spatially in pixels [29,12] and [3,14] moving at equal velocities [6,10] m/s.}
\item{We assumed two targets located near each other spatially in pixels [11,17] [12,18] moving at similar velocities of [8,-12] and [8,-14] m/s, respectively.} 
\end{itemize}
The remaining targets were placed randomly within the scene, with random velocities. We consider two types of clutter important in the moving target imaging: clutter due to large stationary targets and additive Rayleigh distributed stationary clutter embedded in the scene of moving targets. 
Finally, we set $\lambda = 0.2$ for all experiments, which was heuristically chosen for noiseless experiments.

{We note that the assumptions of the moving targets having a linear trajectory while the antennas traverse a circular trajectory may not be valid in practice. However, this configuration is chosen to deconvolve the velocity-estimation effects from potential limited-aperture artifacts.}

\begin{table*}
\centering
\caption{Comparison of Imaging and Scene Parameters}
\label{tab:parameters}
\begin{tabular}{c c c c c c c}
     \hline
     % after \\: \hline or \cline{col1-col2} \cline{col3-col4} ...
     Citation & Method & $f_c$ & $B$ & Resolution & Number of Moving Targets  & Clutter type\\
     \hline
      \cite{onhon2011sar,onhon2017sar} & Sparsity & 15 GHz & 150 MHz & 1m & 5 & Rayleigh distributed\\
      \cite{chen2017refocusing} & Sparsity & 10 GHz & 300 MHz & 0.5m & 1 & Point targets\\
      \cite{8867942} & Sparsity & 1 GHz & 256 MHz & 0.58m & 2 & Point targets\\
      \cite{yang2015strong,li2016moving} & RPCA & 8.85 GHz & 40 MHz & 3.75m & -- & --\\
      \cite{leibovich2019low} & RPCA & 9.6 GHz & 622 MHz & 0.24m & 1 & Point Targets\\
      \cite{Yasin2017,Cetin2018} & RPCA & 15 GHz & 150 MHz & 1m & 2 & Rayleigh distributed\\
     \textbf{Our approach} & \textbf{PSR Recovery} & \textbf{9.45 GHz} & \textbf{50 MHz} & \textbf{3.25m} & \textbf{--} & \textbf{Extended Targets + Rayleigh distributed}\\
     \end{tabular}

\end{table*}
\subsection{Robustness with Respect to Additive Noise in the Data Domain}
In the first set of simulations, we present the reconstruction and velocity estimation moving targets when the received data is contaminated with additive noise. We consider an additive white Gaussian noise process with zero-mean, and define signal-to-noise ratio (SNR) as:
\begin{equation}
 \text{SNR} = 10\log (\frac{\sigma_d}{ \sigma_n}),
\end{equation}
where $\sigma_d$ and $\sigma_n$ are the standard deviations of the received data and noise, respectively. We vary SNR from -20 dB to +20 dB with a step-size of +2 dB, and generate ten realizations. For demonstration purposes, a single realization at SNR = 0 dB is considered. The ground truth scene with antenna trajectories at time $s=0$ is shown in Fig. \ref{fig:ground_truth_traj}. 

\begin{figure}[!h]
\centering
\includegraphics[width = 0.5\textwidth]{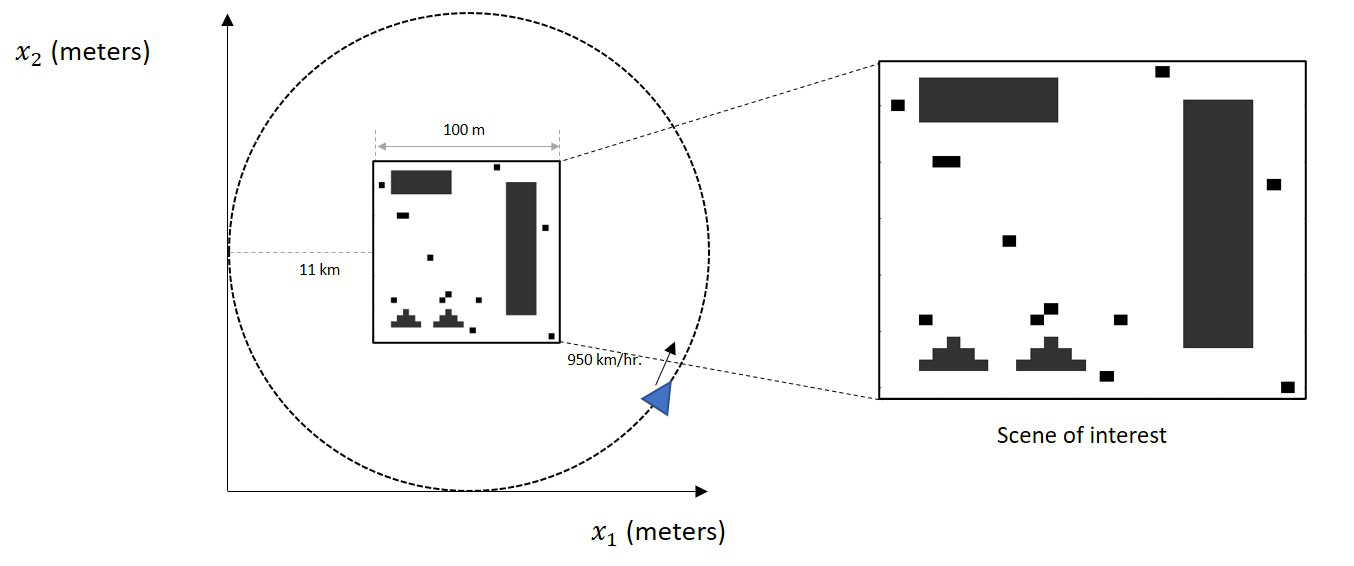}
\caption{Illustration of the simulation set-up for a multiple moving targets and a stationary extended target. The dark region shows the scene of interest, where the gray squares show the positions of the targets. The antennas traverse a circular flight trajectory, where the blue triangle shows the transmitter/receiver pair.}
\label{fig:ground_truth_traj}
\end{figure}
 
To demonstrate algorithm performance, we run PGD, FISTA, and ADMM for 100 iterations. For details of FISTA, we refer the reader to \cite{beck2009fast}. We set $r = t_0 = 1$, where $r$ is a tunable parameter used in both PGD and ADMM, and $t_0$ is a tunable parameter used in the acceleration scheme for FISTA. We consider the $\ell_2$-norm error between the true phase-space reflectivity matrix $\mathbf{Q}$, and the reconstructed phase-space reflectivity matrix $\mathbf{Q}_s + \mathbf{Q}_\nu$. Fig. \ref{fig:loss_plots} shows a plot of the average error for PGD (blue), FISTA (red), and ADMM (yellow) at each iteration. We observe that all three algorithms converge to the same solution, while ADMM converges faster than both PGD and FISTA. Note that in Algorithm \ref{alg:PGD}, the mismatch between $\mathcal{F}^H\mathcal{F}(\mathbf{Q}_s +\mathbf{Q}_\nu)$ and $ \mathcal{F}^H \mathbf{d}$ accumulates in each iteration. While we do not see an impact on our simulated dataset, we keep it in considerations for real-data applications. Meanwhile, we note a reduction in per-iteration runtime when compared to existing methods that utilize generic RPCA due to the efficient implementation of the rank-one constraint.
\begin{figure}[!h]
\centering
\includegraphics[width = 0.35\textwidth]{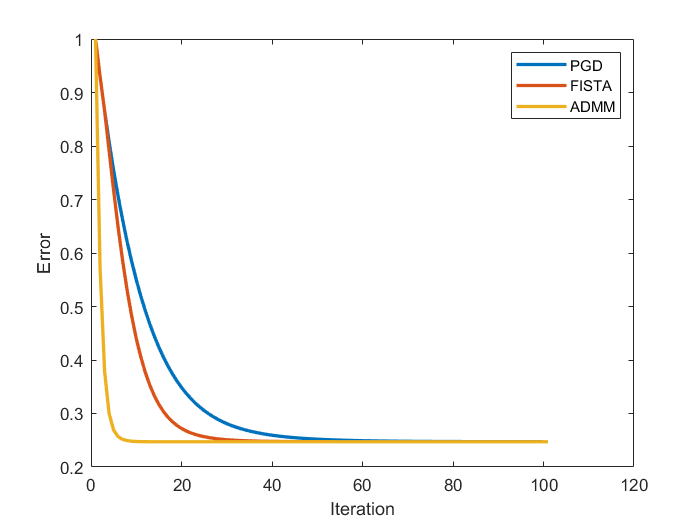}
\caption{$\ell_2$-norm error vs. iterations for PGD (blue) and ADMM (orange).}
\label{fig:loss_plots}

\end{figure}

For visual demonstrations, we display the reconstructed scene using naive backprojection, shown in Figure \ref{fig:bp}. Note that all reconstructed images are displayed in log-scale, with a threshold of -40 dB.
\begin{figure}[!htb]
\centering
\includegraphics[width = 0.24\textwidth]{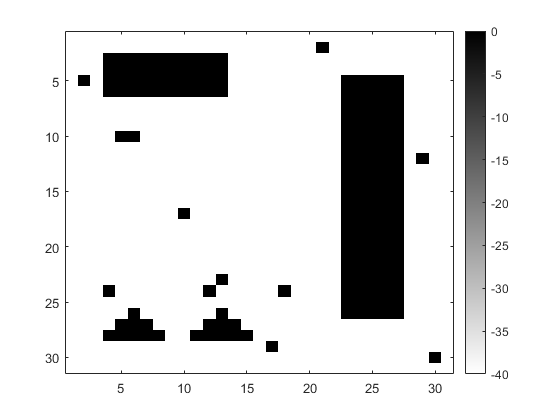}
\includegraphics[width = 0.24\textwidth]{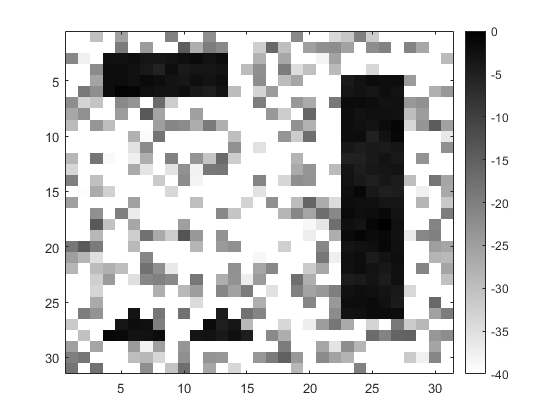}
\caption{The ground truth image (left) and the reflectivity image reconstructed using naive backprojection, $\nu_h = 0$ (right) when SNR = 0 dB.}
\label{fig:bp}
\end{figure}
Clearly, the reflectivities of the stationary targets dominate the scene, and the moving targets cannot be detected. Next, we show the reconstructed moving targets along with the stationary targets using our method in Fig. \ref{fig:moving_single}. The reconstructed moving target scene is formed by superimposing each row of $\mathbf{Q}_\nu$ to form a single image. We see that the moving targets are accurately reconstructed in their correct positions, while the stationary components are also recovered.
\begin{figure}[!htb]
\centering
\includegraphics[width = 0.24\textwidth]{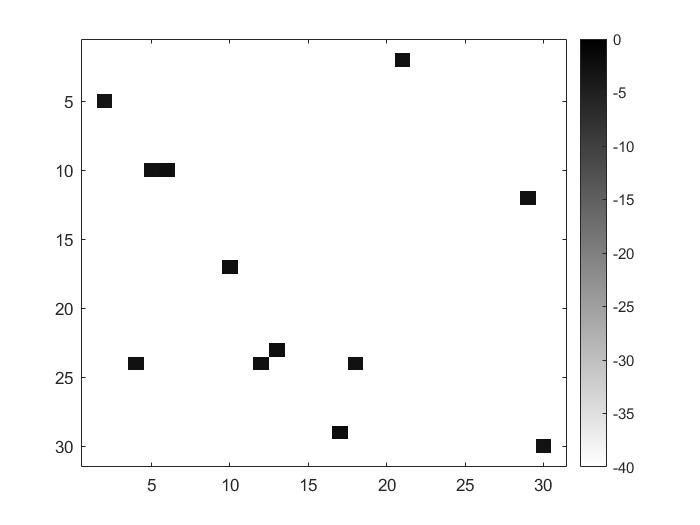}
\includegraphics[width = 0.24\textwidth]{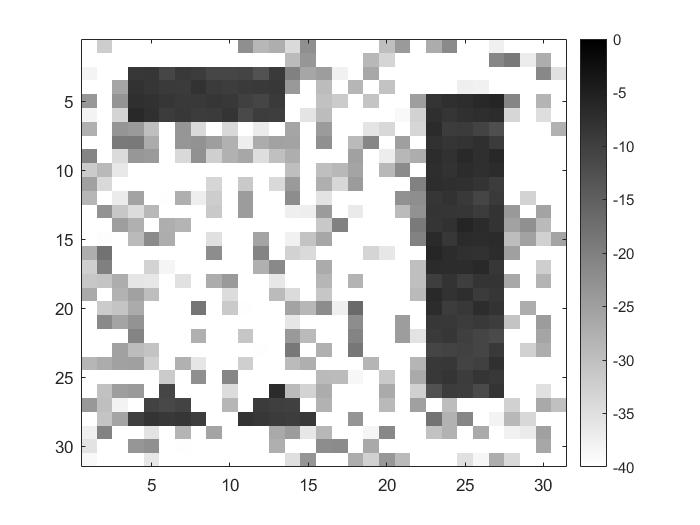}
\includegraphics[width = 0.24\textwidth]{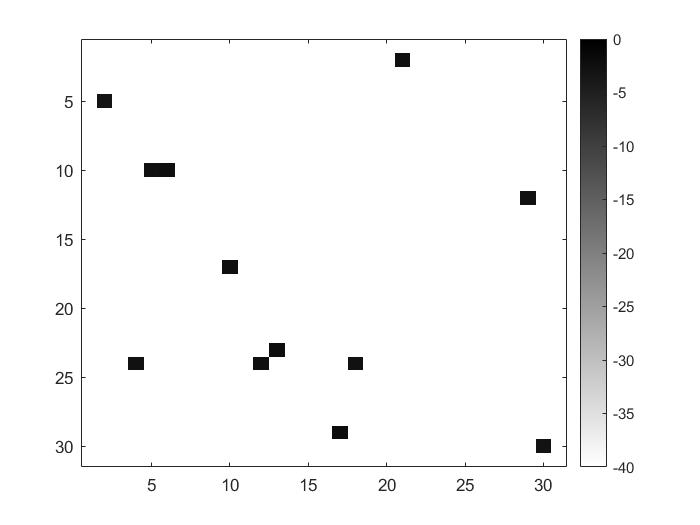}
\includegraphics[width = 0.24\textwidth]{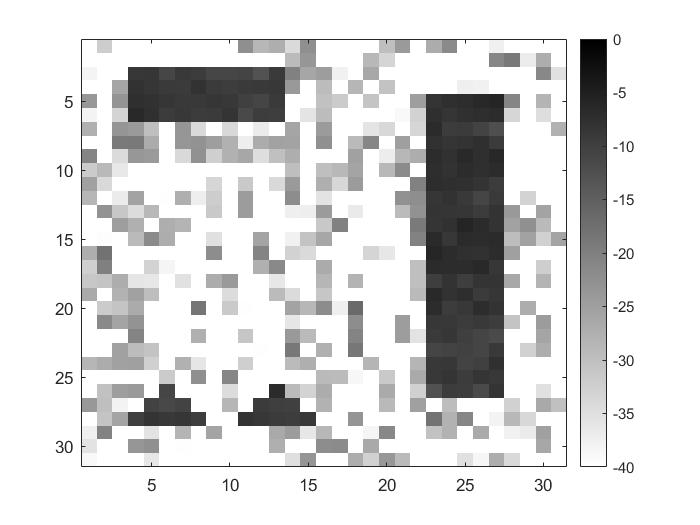}
\caption{The reflectivity image reconstructed using the estimated velocities (left) and the reconstruction of the stationary component (right) for PGD (top) and ADMM (bottom) when SNR = 0 dB.}

\label{fig:moving_single}
\end{figure}
Next, we consider the structural similarity index (SSIM) between the reconstructed moving target scene and the true moving targets in order to provide a quantitative metric for reconstruction. Fig. \ref{fig:ssim_snr} shows the SSIM as SNR varies (blue). The plot indicates accurate recovery of the moving targets beginning at an SNR of -6 dB. We note a graceful degradation of the SSIM curve past this point, indicating a level of robustness with respect to noise.
\begin{figure}[!htb]
\centering
\includegraphics[width = 0.35\textwidth]{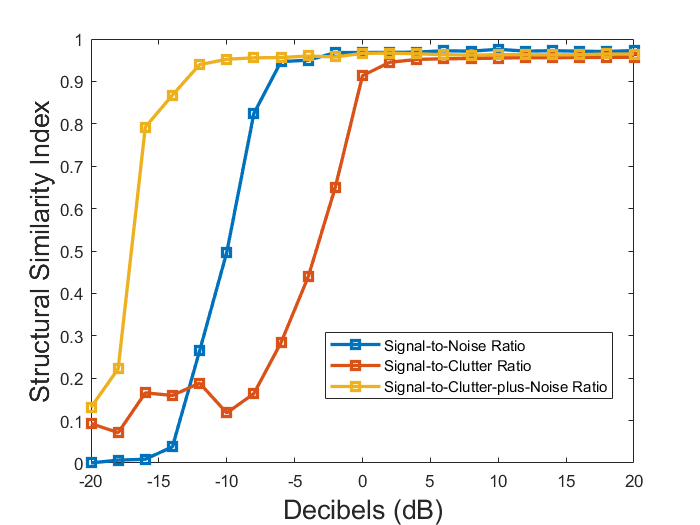}
\caption{This plot shows the resulting SSIM for reconstructed moving target images as SNR (blue), SCR (red) and SCNR (yellow) vary.}
\label{fig:ssim_snr}
\end{figure}
It is also important to consider the amount of false alarms that arise. Since we fix $\lambda$, we do not expect to threshold out the moving targets as SNR decreases, but we do expect the number of false alarms to increase. For this reason, we introduce the positive predictive value (PPV), which is the ratio of true positives to the total number of detections:
\begin{align}\label{eq:PPV}
\textrm{PPV} &= \frac{true\ positives}{true\ positives + false\ positives}
\end{align}
Fig. \ref{fig:prob_det} shows the PPV as SNR (blue) varies. We compute the PPV over the entire PSR matrix, noting that a target could either be reconstructed in the incorrect location, with the incorrect velocity estimate, or both. A true positive is defined as detecting a target in its correct location spatially with the correct velocity, while a false positive occurs with either an incorrect location or incorrect velocity. We see that the algorithm produces no false alarms at an SNR of -2 dB. This indicates that there are many low-intensity false alarms even when the moving targets are reconstructed at low SNR values. This can be addressed through hyperparameter tuning.
\begin{figure}[!h]
\centering
\includegraphics[width = 0.35\textwidth]{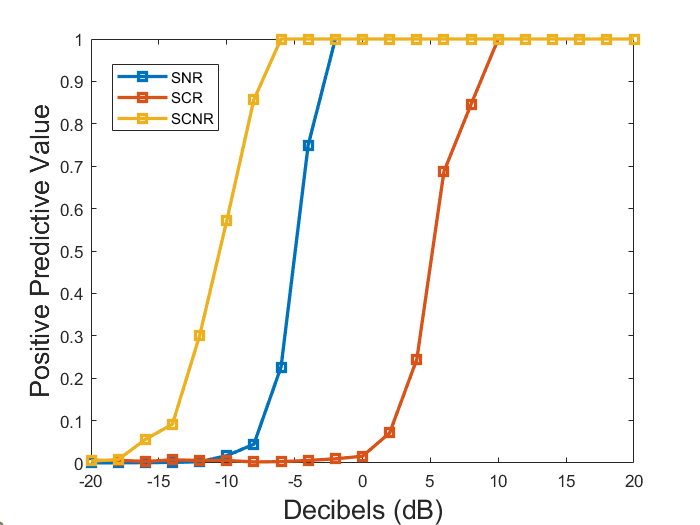}
\caption{This plot shows the resulting PPV at various levels of SNR (blue), SCR (red), and SCNR (yellow).}
\label{fig:prob_det}
\end{figure}
\subsection{Robustness with Respect to Clutter}

The second set of simulations demonstrates the performance of the algorithms with respect to clutter. Clutter was added to the scene using a Rayleigh distribution for the amplitude, a commonly used statistical model for clutter in radar data \cite{shnidman1999generalized}. We define the signal-to-clutter ratio as 
\begin{equation}
\text{SCR} = 10\log (\frac{\sigma_f}{\sigma_b}),
\end{equation} where $\sigma_f$ and $\sigma_b$ are the standard deviations of the moving target foreground and clutter background intensity values, respectively. Different SCR scenarios were simulated by keeping the moving target amplitudes fixed and varying the standard deviation of the background. We varied the SCR from -20 dB to +20 dB with a step-size of +2 dB. Ten realizations of clutter were generated. For visual demonstrations, we consider a single realization of SCR = 0 dB shown in Fig. \ref{fig:scr_traj}. The reconstruction results are shown in Fig. \ref{fig:scr_recon}. We see that the moving targets are focused in the correct positions, while the stationary clutter is simultaneously reconstructed.

\begin{figure}[h]
\centering
\includegraphics[width = 0.5\textwidth]{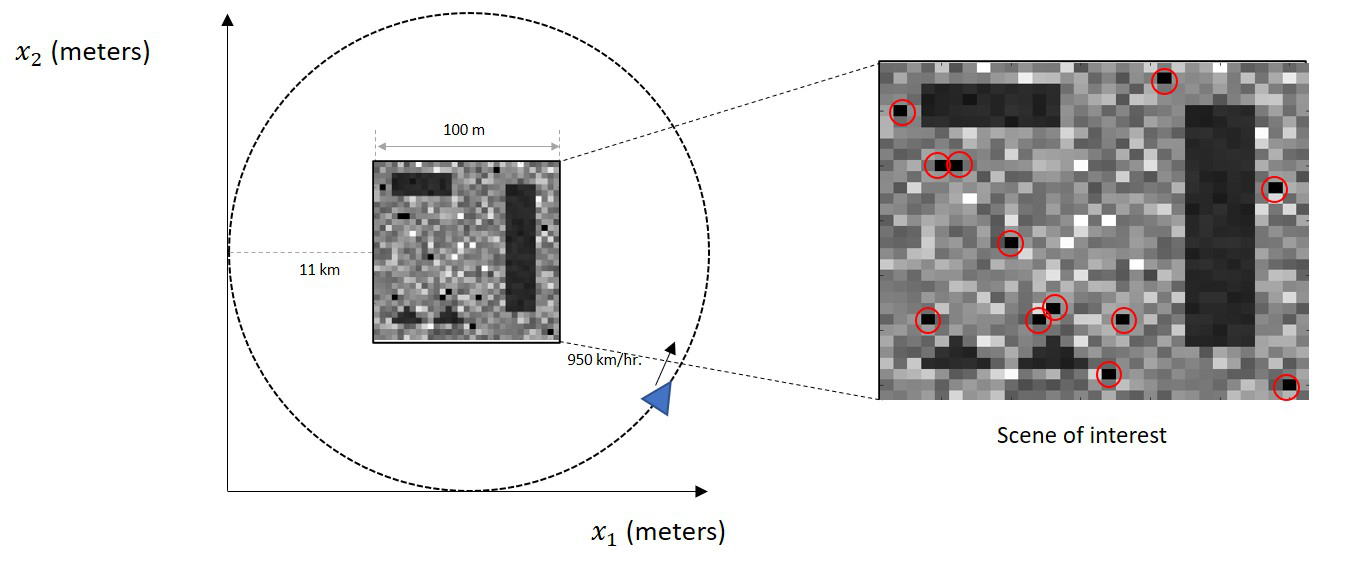}
\caption{Illustration of the simulation set-up for a multiple moving targets at SCR = 0 dB. The dark region shows the scene of interest, where the red circles show the positions of the targets embedded in clutter. The antennas traverse a circular flight trajectory, where the blue triangle shows the transmitter/receiver pair.}
\label{fig:scr_traj}
\end{figure}

\begin{figure}[!htb]
\centering
\includegraphics[width = 0.24\textwidth]{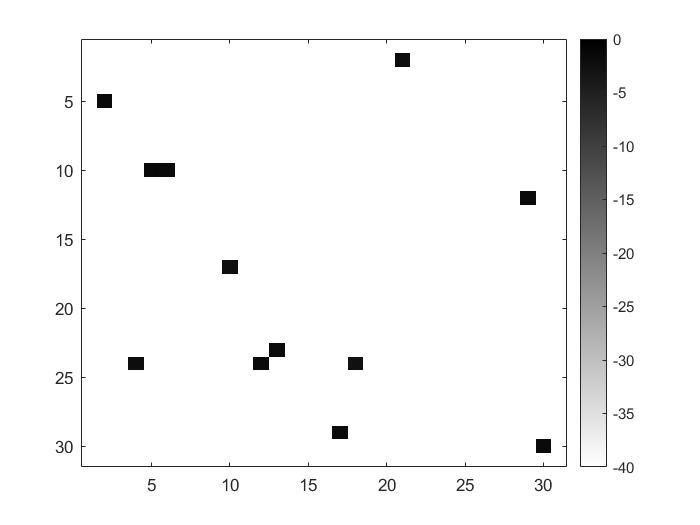}
\includegraphics[width = 0.24\textwidth]{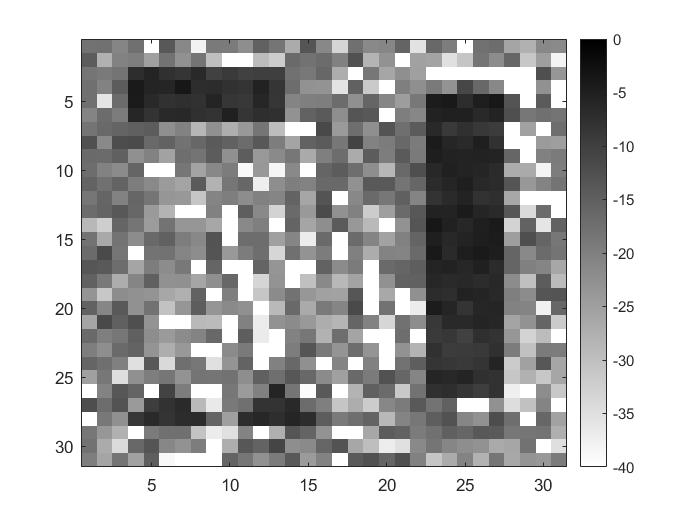}
\caption{The reflectivity image reconstructed using the estimated velocities (left) and the reconstruction of the stationary component (right) when SCR = 0 dB.}

\label{fig:scr_recon}
\end{figure}
Fig. \ref{fig:ssim_snr} and Fig. \ref{fig:prob_det} show the plots of the SSIM and PPV as SCR varies (red), portraying accurate reconstruction at higher SCR levels. We see a decline in performance after 0 dB, indicating that the clutter amplitude is too strong for the algorithm to detect the targets. We see that the algorithm also requires higher SCR levels to detect the moving targets with no false alarms, but the curve indicates a degree of robustness with respect to clutter, and can be addressed through hyperparameter tuning. Our simulations show this method works particularly well in low-SCR environments in reference to the reported results of the sub-aperture RPCA method in \cite{Cetin2018}, which exhibits similar performance at an SCR of +10 dB. In addition, \cite{Cetin2018} requires a sparsity-driven focusing algorithms once the moving targets are separated from the stationary scene, while our proposed method does not. 
\subsection{Robustness with Respect to Noise and Clutter}
\textcolor{black}
{In the third set of simulations, we explore the effects of having both noise and clutter on performance of reconstructing targets and estimating velocities. Here, we fix the clutter background to 0 dB and define signal-to-clutter-plus-noise ratio as 
\begin{equation}
\text{SCNR} = \frac{\sigma_d}{\sigma_c + \sigma_n},
\end{equation}
where $\sigma_d$, $\sigma_c$, and $\sigma_n$ are the standard deviations of the data received by the moving targets, the data received by the clutter background, and the noise, respectively. We varied the SCNR from -20 dB to +20 dB with a step-size of +2 dB by varying $\sigma_n$. Fig. \ref{fig:scnr_recon} shows the reconstruction of a single realization when SCNR = 0 dB. We plot the SSIM as we vary SCNR (yellow) in Fig. \ref{fig:ssim_snr}, which portrays accurate reconstruction of targets at -10 dB, with performance declining at SCNR levels lower than that. Similar to previous observations, we see that the the algorithm requires higher levels of SCNR to detect moving targets with no false alarms, shown in Fig. \ref{fig:prob_det}.}
\begin{figure}[!htb]
\centering
\includegraphics[width = 0.24\textwidth]{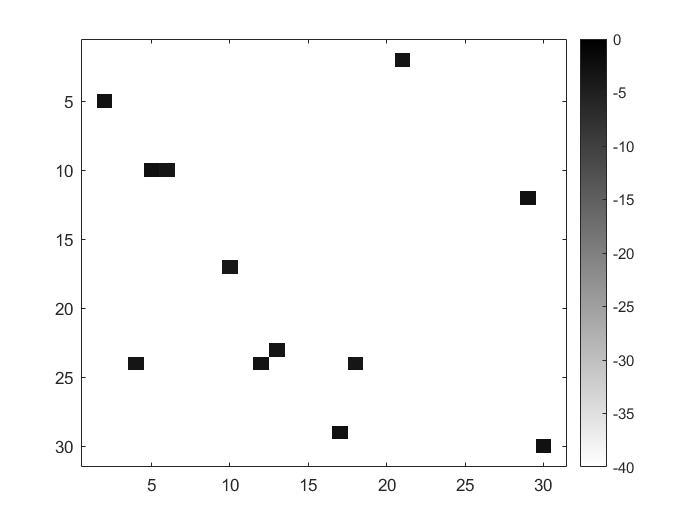}
\includegraphics[width = 0.24\textwidth]{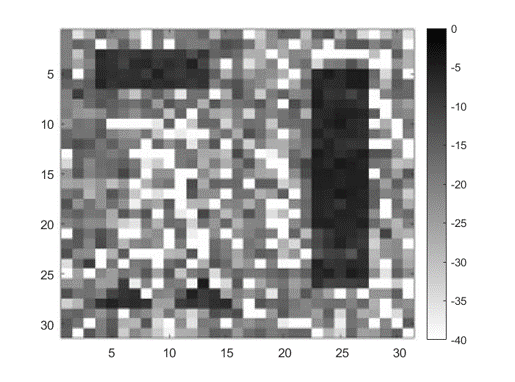}
\caption{The reflectivity image reconstructed using the estimated velocities (left) and the reconstruction of the stationary component (right) when SCNR = 0 dB.}
\label{fig:scnr_recon}
\end{figure}
\subsection{Dense Moving Target Imaging}
Finally, we discuss the effects of observing many moving targets within the scene of interest. In this set of experiments, we set the number of moving targets to 250, such that they occupy approximately a quarter of the scene. For all experiments, we consider scenarios in which SNR = 0 dB.

We note that when considering a cardinality constraint, it is known that the number of false alarms is heavily dependent on the choice of $k$. In this case, $k$ corresponds to the number of moving targets known a priori, which is often difficult to determine in practice.  Meanwhile, in the convex case, it is known that increasing $\lambda$ promotes a sparser solution. Thus, we plot the number of false alarms as we vary both $k$ and $\lambda$, shown in Fig. \ref{fig:varied_param}.

\begin{figure}[!htb]
\centering
\includegraphics[width = 0.35\textwidth]{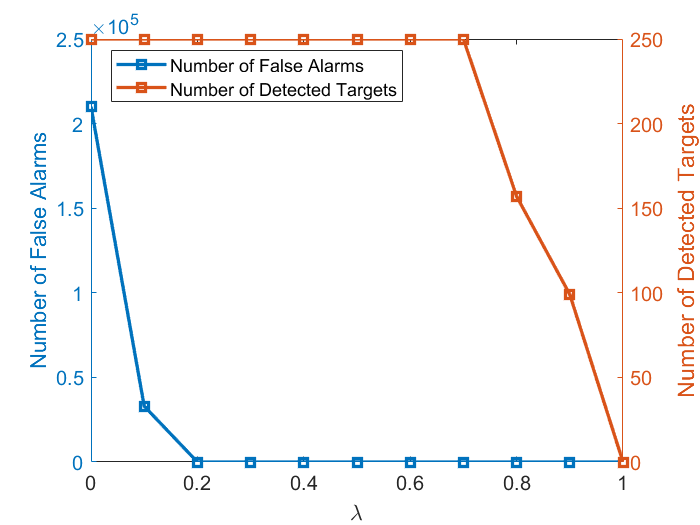}
\includegraphics[width = 0.35	\textwidth]{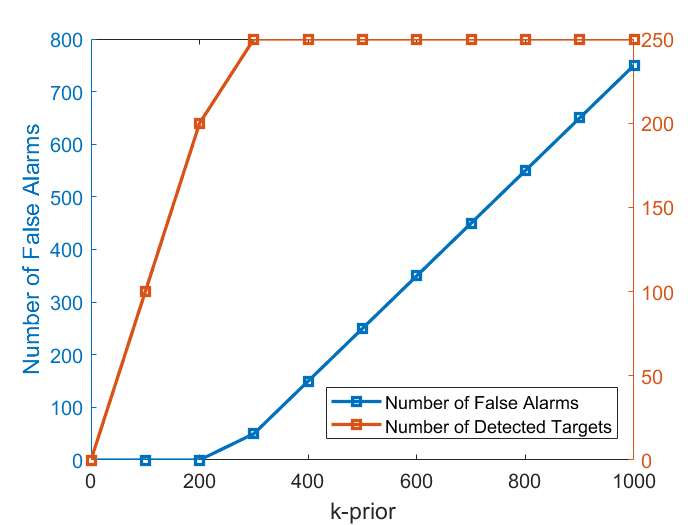}
\caption{Number of false alarms (blue) and detected targets (orange) as $\lambda$ is varied (top) and $k$-prior is varied (bottom).}
\label{fig:varied_param}
\end{figure}

It is clear that the performance of Algorithm \ref{alg:non-convex alg} degrades when the number of moving targets is not exactly known. If the estimate is too low, then the algorithm cannot detect the true moving targets, and if the estimate is too high, then many false alarms will be present. As expected, this algorithm has similar limitations to methods such as \cite{yang2015strong}, which shows that the percentage of correct detections converges to an upper-limit of about 80\%, while the percentage of false alarms increases linearly with respect to the threshold $k$. However, our observations do show a speedup in convergence due to the implementation of the rank-constraint.

On the other hand, when considering Algorithm \ref{alg:PGD}, we observe that the false alarms can be threshold out by increasing $\lambda$. Typically, there is a tradeoff between the density of the scene and the $\lambda$ value, since a higher $\lambda$ will promote sparsity. However, even though the scene is dense, $\mathbf{Q}_\nu$ remains sparse as the scene is mapped from $\mathbb{R}^N \to \mathbb{R}^{M\times N}$. This is in contrast to methods that formulate the problem over the data matrix, as the received echoes from multiple moving targets may not satisfy the high level of sparsity assumptions required in \cite{Borcea2012, leibovich2019low}.  Comparing Algorithms \ref{alg:PGD} and \ref{alg:non-convex alg}, we see that there is a large acceptable range of $\lambda$ values that detect all moving targets while suppressing false alarms. This addresses the problem of requiring exact knowledge of the number of moving targets, without the need of multiple receivers or additional pre-processing methods such as \cite{li2016moving}.
\section{Conclusion}\label{Conclusion}
In this paper, we present a novel methodology for ground moving target imaging and velocity estimation based on the phase-space reflectivities. We exploit the structure of the phase-space reflectivity matrix and formulate the problem as the recovery of a rank-one stationary component and a sparse component associated with moving targets. We show that these two components lie in orthogonal subspaces, where their support sets are disjoint from each other. Our method is capable of imaging moving targets regardless of their speed or direction, as there is no prior assumption about target motion parameters other than assuming constant velocity. Furthermore, the phase-space reflectivity matrix can be obtained efficiently through fast backprojection algorithms.

We derive two algorithms to solve the moving target imaging problem, focusing on convex methods, as well as a non-convex instance that assumes a priori knowledge of the number of moving targets in the scene. We show that having set the hypothesized velocities a priori, we directly enforce the rank constraint by a known support, keeping the problem convex. Compared to generic RPCA methods found in literature, our formulation leads to a more computationally efficient implementation, reducing per-iteration complexity by an order of magnitude. Meanwhile, compared to sparsity-driven methods, we do not need any clutter suppression techniques. Additionally, since the supports of the stationary and moving target components are fully decoupled, we do not rely on having to distinguish the $\ell_1-$ and nuclear-norms to maintain separability. 

We use numerical simulations to demonstrate the capability of our method to reconstruct focused images of the moving targets and estimate their velocities. We show that the moving targets can be reconstructed when the targets are near large stationary objects or embedded in stationary clutter. Simulations show that the algorithms are robust at reconstructing the moving targets in the presence of both additive noise and clutter. Furthermore, we show that introducing prior knowledge of the number of moving targets can help mitigate the number of false alarms, but the performance heavily depends on this knowledge which is difficult to determine in practice.

In future work, many aspects of performance analysis can be explored. The algorithms assume that we have an accurate set of hypothesized velocities for the moving targets, but it would be useful to consider an off-the-grid approach to estimate velocities in the continuous-domain. Furthermore, a sample complexity analysis on the forward would be useful in determining the sensitivity of the velocity estimation of this approach. Meanwhile, an analysis that studies super-resolution capabilities of the algorithms can be performed. Although we focus on verifying our algorithm through numerical simulations, it would be beneficial to implement our method on real data. This would provide insight on the practicality of our algorithms in modern radar scenarios. Finally, we note this work can be extended through the concept unrolling iterative algorithms into a deep network architecture \cite{solomon2019deep}. This application would be particularly useful in passive SAR, as the location of transmitters and the transmitted waveforms are unknown. Meanwhile, for active SAR, deep learning provides the ability to implement the inverse operators in the update equations, as our iterative scheme relies on simplifying assumptions due to computational concerns. We leave the deep learning direction for our upcoming studies.

\bibliographystyle{IEEEtran}

%Please number citations consecutively within brackets \cite{b1}. The 
%sentence punctuation follows the bracket \cite{b2}. Refer simply to the reference 
%number, as in \cite{b3}---do not use ``Ref. \cite{b3}'' or ``reference \cite{b3}'' except at 
%the beginning of a sentence: ``Reference %\cite{b3} was the first $\ldots$''

%\begin{thebibliography}{00}
%\bibitem{b1} G. Eason, B. Noble, and I. N. Sneddon, ``On certain integrals of Lipschitz-Hankel type involving products of Bessel functions,'' Phil. Trans. Roy. Soc. London, vol. A247, pp. 529--551, April 1955.
%\bibitem{b2} J. Clerk Maxwell, A Treatise on Electricity and Magnetism, 3rd ed., vol. 2. Oxford: Clarendon, 1892, pp.68--73.
%\bibitem{b3} I. S. Jacobs and C. P. Bean, ``Fine particles, thin films and exchange anisotropy,'' in Magnetism, vol. III, G. T. Rado and H. Suhl, Eds. New York: Academic, 1963, pp. 271--350.
%\bibitem{b4} K. Elissa, ``Title of paper if known,'' unpublished.
%\bibitem{b5} R. Nicole, ``Title of paper with only first word capitalized,'' J. Name Stand. Abbrev., in press.
%\bibitem{b6} Y. Yorozu, M. Hirano, K. Oka, and Y. Tagawa, ``Electron spectroscopy studies on magneto-optical media and plastic substrate interface,'' IEEE Transl. J. Magn. Japan, vol. 2, pp. 740--741, August 1987 [Digests 9th Annual Conf. Magnetics Japan, p. 301, 1982].
%\bibitem{b7} M. Young, The Technical Writer's Handbook. Mill Valley, CA: University Science, 1989.
%\end{thebibliography}
\bibliography{references}
\begin{IEEEbiography}[{\includegraphics[width=1in,height=1.25in,clip,keepaspectratio]{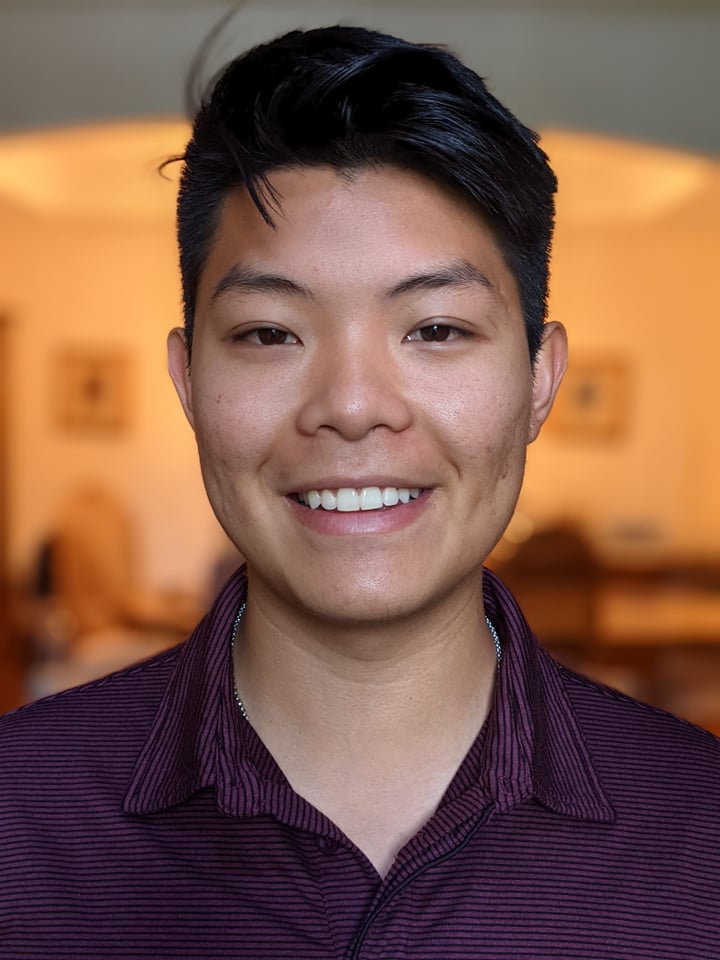}}]{Sean Thammakhoune}
received his B.Sc. degrees in electrical engineering and computer and systems engineering from Rensselaer Polytechnic Institute (RPI) in Troy, NY, USA in May 2018, where he is currently working toward the Ph.D. degree in electrical engineering with the Computational Imaging Group. His research interests include optimization methods for inverse problems and imaging in radar. \end{IEEEbiography}

% if you will not have a photo at all:
\begin{IEEEbiography}[{\includegraphics[width=1in,height=1.25in,clip,keepaspectratio]{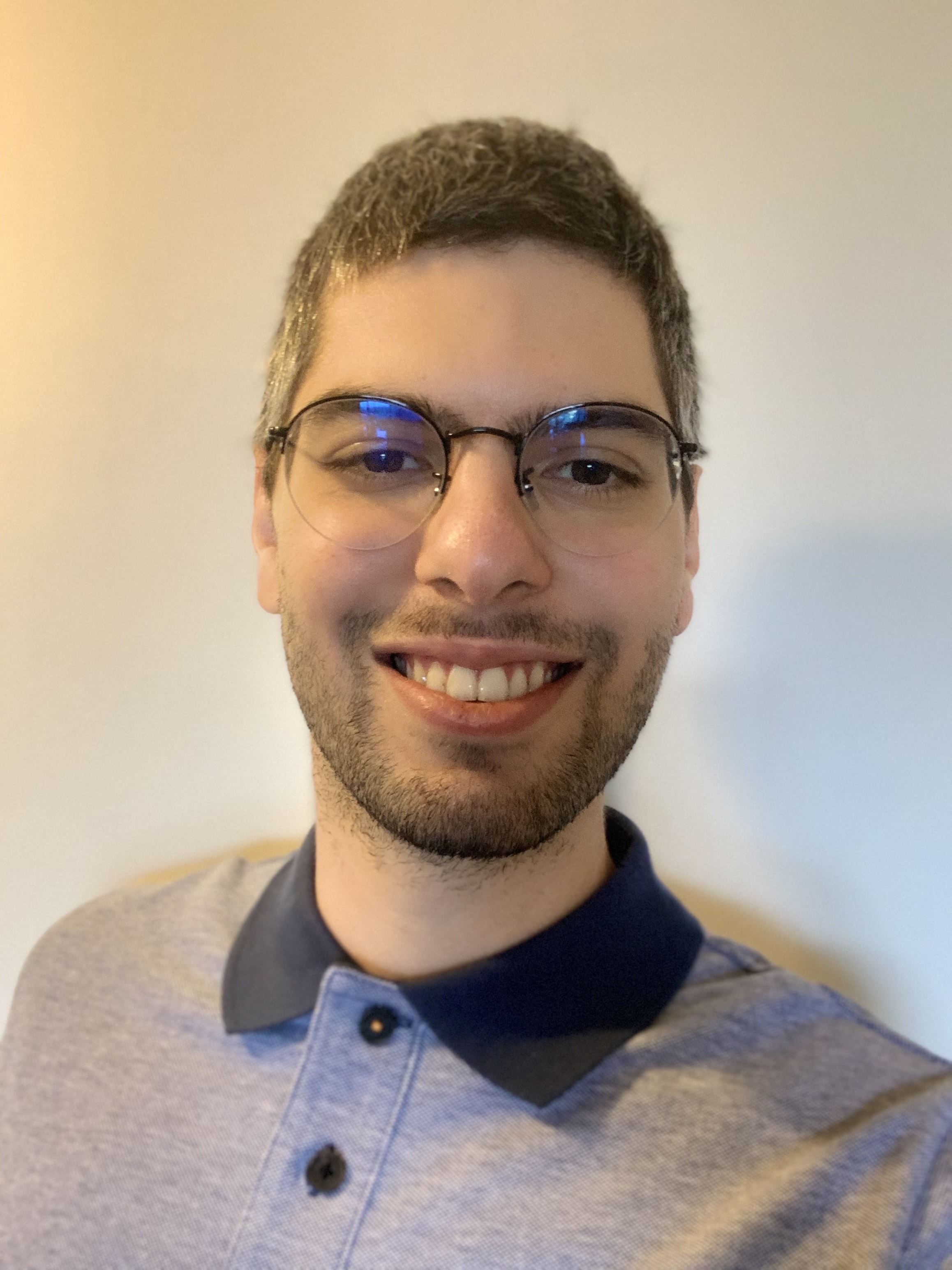}}]{Bariscan Yonel}
(Member, IEEE) received his B.Sc. degree in electrical engineering from Koc University in Istanbul, Turkey in Jun. 2015, and his Ph.D. in electrical engineering from Rensselaer Polytechnic Institute (RPI) in Troy, NY, USA in Dec. 2020, where he is currently a postdoctoral research associate with the Computational Imaging Group. His work focuses on theoretical guarantees and practical limitations for solving quadratic equations in high dimensional inference and wave-based imaging problems, using low rank matrix recovery theory and computationally efficient non-convex algorithms. Research interests include applications and performance analysis of machine learning, compressed sensing and optimization methods for inverse problems in imaging and signal processing.\end{IEEEbiography}

% insert where needed to balance the two columns on the last page with
% biographies
%\newpage

\begin{IEEEbiography}[{\includegraphics[width=1in,height=1.25in,clip,keepaspectratio]{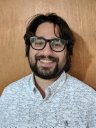}}]{Eric Mason}
received his B.S. degree in electrical engineering from Manhattan College, Bronx, NY, USA, in 2012 and the Ph.D. in electrical engineering from Rensselaer Polytechnic Institute, Troy, NY, USA, in 2017. In summer 2017, he joined the U.S. Naval Research Laboratory, Washington, D.C., USA, where he currently works in the Tactical Electronic Warfare Division performing basic and applied research in the areas of signal processing, optimization and machine learning. His research interests include optimization methods for inverse problems in radar and radio-frequency application of machine learning.\end{IEEEbiography}
 
\begin{IEEEbiography}[{\includegraphics[width=1in,height=1.25in,clip,keepaspectratio]{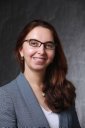}}]{Birsen Yazici}
(Fellow Member, IEEE) received the B.S. degrees in electrical engineering and mathematics, in 1988, from Bogazici University, Istanbul Turkey, and M.S. and Ph.D. degrees in mathematics and electrical engineering both from Purdue University, West Lafayette, IN, in 1990 and 1994, respectively. From September 1994 until 2000, she was a Research Engineer at General Electric Company Global Research Center, Schenectady NY. During her tenure in industry, she worked on radar, transportation, industrial and medical imaging systems. From
2001 to June 2003, she was an Assistant Professor at Drexel University, Electrical and Computer Engineering Department. In 2003, she joined Rensselaer Polytechnic Institute where she is currently a Full Professor in the Department of Electrical, Computer and Systems Engineering and in the Department of Biomedical Engineering. Prof. Yazıcı’s research interests span the areas of statistical signal processing, inverse problems in imaging, image reconstruction, biomedical optics, radar and X-ray imaging. She served as an associate editor for the IEEE TRANSACTIONS ON IMAGE PROCESSING from 2008 to 2012, IEEE TRANSACTIONS ON GEOSCIENCES AND REMOTE SENSING from 2014 to 2018, for SIAM Journal on Imaging Science from 2010 to 2014, and for IEEE TRANSACTIONS ON COMPUTATIONAL IMAGING, from 2017 to 2020.  She is currently a Distinguished Lecturer of the IEEE Aerospace and Electronics Systems Society. She is the recipient of the Rensselaer Polytechnic Institute 2007 and 2013 School of Engineering Research Excellence awards. She holds 11 US patents.\end{IEEEbiography}

\begin{IEEEbiography}[{\includegraphics[width=1in,height=1.25in,clip,keepaspectratio]{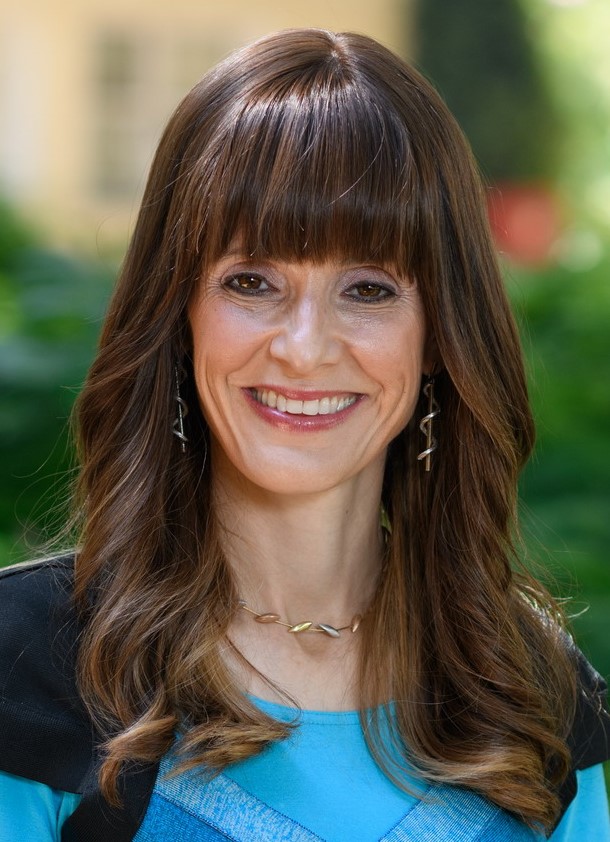}}]{Yonina C. Eldar}
(Fellow, IEEE) received the B.Sc. degree in Physics in 1995 and the B.Sc. degree in Electrical Engineering in 1996 both from Tel-Aviv University (TAU), Tel-Aviv, Israel, and the Ph.D. degree in Electrical Engineering and Computer Science in 2002 from the Massachusetts Institute of Technology (MIT), Cambridge. She is currently a Professor in the Department of Mathematics and Computer Science, Weizmann Institute of Science, Rehovot, Israel. She was previously a Professor in the Department of Electrical Engineering at the Technion. She is also a Visiting Professor at MIT, a Visiting Scientist at the Broad Institute, and an Adjunct Professor at Duke University and was a Visiting Professor at Stanford. She is a member of the Israel Academy of Sciences and Humanities (elected 2017), an IEEE Fellow and a EURASIP Fellow. Her research interests are in the broad areas of statistical signal processing, sampling theory and compressed sensing, learning and optimization methods, and their applications to biology, medical imaging and optics.

Dr. Eldar has received many awards for excellence in research and teaching, including the IEEE Signal Processing Society Technical Achievement Award (2013), the IEEE/AESS Fred Nathanson Memorial Radar Award (2014), and the IEEE Kiyo Tomiyasu Award (2016). She was a Horev Fellow of the Leaders in Science and Technology program at the Technion and an Alon Fellow. She received the Michael Bruno Memorial Award from the Rothschild Foundation, the Weizmann Prize for Exact Sciences, the Wolf Foundation Krill Prize for Excellence in Scientific Research, the Henry Taub Prize for Excellence in Research (twice), the Hershel Rich Innovation Award (three times), the Award for Women with Distinguished Contributions, the Andre and Bella Meyer Lectureship, the Career Development Chair at the Technion, the Muriel $\&$ David Jacknow Award for Excellence in Teaching, and the Technion’s Award for Excellence in Teaching (two times).  She received several best paper awards and best demo awards together with her research students and colleagues including the SIAM outstanding Paper Prize, the UFFC Outstanding Paper Award, the Signal Processing Society Best Paper Award and the IET Circuits, Devices and Systems Premium Award, was selected as one of the 50 most influential women in Israel and in Asia, and is a highly cited researcher.

She was a member of the Young Israel Academy of Science and Humanities and the Israel Committee for Higher Education. She is the Editor in Chief of Foundations and Trends in Signal Processing, a member of the IEEE Sensor Array and Multichannel Technical Committee and serves on several other IEEE committees. In the past, she was a Signal Processing Society Distinguished Lecturer, member of the IEEE Signal Processing Theory and Methods and Bio Imaging Signal Processing technical committees, and served as an associate editor for the IEEE Transactions On Signal Processing, the EURASIP Journal of Signal Processing, the SIAM Journal on Matrix Analysis and Applications, and the SIAM Journal on Imaging Sciences. She was Co-Chair and Technical Co-Chair of several international conferences and workshops. She is author of the book "Sampling Theory: Beyond Bandlimited Systems" and co-author of four other books published by Cambridge University Press.
\end{IEEEbiography}

\end{document}